%% file: main.tex
\pdfoutput=1

\documentclass[12pt,a4paper]{article}

\usepackage{ifthen} 
\newboolean{pdflatex}
\setboolean{pdflatex}{true} 

\newboolean{articletitles}
\setboolean{articletitles}{true} 

\newboolean{uprightparticles}
\setboolean{uprightparticles}{false} 

\newboolean{inbibliography}
\setboolean{inbibliography}{false} 

\input{preamble}
\usepackage{longtable} 
\input{lhcb-symbols-def}

\begin{document}

\renewcommand{\thefootnote}{\fnsymbol{footnote}}
\setcounter{footnote}{1}

\input{title-LHCb-PAPER}


\renewcommand{\thefootnote}{\arabic{footnote}}
\setcounter{footnote}{0}



\pagestyle{plain} 
\setcounter{page}{1}
\pagenumbering{arabic}


%

\input{Introduction}

\input{Detector}

\input{Selection}

\input{Backgrounds}

\input{MassFit}

\input{Angles_Acceptance_Backgrounds}

\input{FullFit}

\input{Summary}


\input{acknowledgements}



\addcontentsline{toc}{section}{References}
\setboolean{inbibliography}{true}
\bibliographystyle{LHCb}
\bibliography{main,LHCb-PAPER,LHCb-CONF,LHCb-DP,LHCb-TDR}

\newpage


\input{LHCb_HD_authorlist_2014-11-18}

\end{document}

%% file: preamble.tex
\usepackage{microtype}
\usepackage{lineno}  
\usepackage{xspace} 
\usepackage{caption} 

\usepackage{graphicx}  
\usepackage{color}
\usepackage{colortbl}
\graphicspath{{./figs/}} 

\usepackage{amsmath} 
\usepackage{amssymb}
\usepackage{amsfonts}
\usepackage{upgreek} 

\newcommand*\patchAmsMathEnvironmentForLineno[1]{%
\expandafter\let\csname old#1\expandafter\endcsname\csname #1\endcsname
\expandafter\let\csname oldend#1\expandafter\endcsname\csname
end#1\endcsname
 \renewenvironment{#1}%
   {\linenomath\csname old#1\endcsname}%
   {\csname oldend#1\endcsname\endlinenomath}%
}
\newcommand*\patchBothAmsMathEnvironmentsForLineno[1]{%
  \patchAmsMathEnvironmentForLineno{#1}%
  \patchAmsMathEnvironmentForLineno{#1*}%
}
\AtBeginDocument{%
\patchBothAmsMathEnvironmentsForLineno{equation}%
\patchBothAmsMathEnvironmentsForLineno{align}%
\patchBothAmsMathEnvironmentsForLineno{flalign}%
\patchBothAmsMathEnvironmentsForLineno{alignat}%
\patchBothAmsMathEnvironmentsForLineno{gather}%
\patchBothAmsMathEnvironmentsForLineno{multline}%
\patchBothAmsMathEnvironmentsForLineno{eqnarray}%
}

\usepackage{hyperref}    
\usepackage[all]{hypcap} 
\usepackage{cite} 
\usepackage{mciteplus}

%% file: lhcb-symbols-def.tex
\def\BdToKstPizGamma  {\decay{\Bd}{\Kstarz(\ra \KS \piz) \g}}
\def\meeKstar {\ensuremath{ m(K^+\pi^- \epem)}}
\def\BdKstGamToee {\decay{\Bd}{\Kstarz \g_{\epem}}}
\def\KstToKPi    {\decay{\Kstarz}{K^+\pi^-}}
\def\BdKstV    {\decay{\Bd}{\Kstarz V(\ra \epem)}}
\def\BdKstEta    {\decay{\Bd}{\Kstarz\Peta}}
\def\BdKstPiz    {\decay{\Bd}{\Kstarz\piz}}

\def\BdToeeKst    {\decay{\Bd}{\Kstarz\ep\en}}
\def\BsToPhiee    {\decay{\Bs}{\phi\ep\en}}
\def\LbToeeL    {\decay{\Lb}{\Pp K^- \ep\en}}
\def\BdToJPsieeKst  {\decay{\Bd}{\jpsi(\epem)\Kstarz}}
\def\ATD     {\ensuremath{A_{\mathrm{T}}^{(2)}}\xspace}
\def\ATIm     {\ensuremath{A_{\mathrm{T}}^{\mathrm{Im}}}\xspace}
\def\ATRe     {\ensuremath{A_{\mathrm{T}}^{\mathrm{Re}}}\xspace}
\def\FL       {\ensuremath{F_{\mathrm{L}}}\xspace}
\def\ATDPH    {\ensuremath{A_{\mathrm{T,PR}}^{(2)}}\xspace}
\def\ATImPH     {\ensuremath{A_{\mathrm{T,PR}}^{\mathrm{Im}}}\xspace}
\def\ATRePH     {\ensuremath{A_{\mathrm{T,PR}}^{\mathrm{Re}}}\xspace}
\def\FLPH      {\ensuremath{F_{\mathrm{L,PR}}}\xspace}
\def\ATDKG    {\ensuremath{A_{\mathrm{T,\Kstarz \gamma}}^{(2)}}\xspace}
\def\ATImKG     {\ensuremath{A_{\mathrm{T,\Kstarz \gamma}}^{\mathrm{Im}}}\xspace}
\def\ATReKG     {\ensuremath{A_{\mathrm{T,\Kstarz \gamma}}^{\mathrm{Re}}}\xspace}
\def\FLKG       {\ensuremath{F_{\mathrm{L,\Kstarz \gamma}}}\xspace}

\def\ctl       {\ensuremath{\cos{\theta_\ell}}\xspace}
\def\ctk       {\ensuremath{\cos{\theta_K}}\xspace}
\def\phit       {\ensuremath{\tilde\phi}\xspace}
\newcommand{\gevgevcccc}{\ensuremath{{\mathrm{\,Ge\kern -0.1em V^2\!/}c^4}}\xspace}

\def\lhcb {\mbox{LHCb}\xspace}

\def\MagUp {\mbox{\em Mag\kern -0.05em Up}\xspace}

\ifthenelse{\boolean{uprightparticles}}%
{
 
 \def\Pgamma      {\ensuremath{\upgamma}\xspace}

 \def\Peta        {\ensuremath{\upeta}\xspace}

 \def\Pmu         {\ensuremath{\upmu}\xspace}

 \def\Ppi         {\ensuremath{\uppi}\xspace}                 
                  
 \def\Prho        {\ensuremath{\uprho}\xspace}

 \def\Pphi        {\ensuremath{\upphi}\xspace}

 \def\Ppsi        {\ensuremath{\uppsi}\xspace}                 
 \def\Pomega      {\ensuremath{\upomega}\xspace}                 

 \def\PDelta      {\ensuremath{\Delta}\xspace}                 
 \def\PXi      {\ensuremath{\Xi}\xspace}                 
 \def\PLambda      {\ensuremath{\Lambda}\xspace}                 
 \def\PSigma      {\ensuremath{\Sigma}\xspace}                 
 \def\POmega      {\ensuremath{\Omega}\xspace}                 
 \def\PUpsilon      {\ensuremath{\Upsilon}\xspace}                 
 
 \def\PB      {\ensuremath{\mathrm{B}}\xspace}                 
                  
 \def\PD      {\ensuremath{\mathrm{D}}\xspace}

 \def\PJ      {\ensuremath{\mathrm{J}}\xspace}                 
 \def\PK      {\ensuremath{\mathrm{K}}\xspace}

 \def\Pb      {\ensuremath{\mathrm{b}}\xspace}                 
 \def\Pc      {\ensuremath{\mathrm{c}}\xspace}                 
                  
 \def\Pe      {\ensuremath{\mathrm{e}}\xspace}

 \def\Pi      {\ensuremath{\mathrm{i}}\xspace}

 \def\Pp      {\ensuremath{\mathrm{p}}\xspace}

 \def\Ps      {\ensuremath{\mathrm{s}}\xspace}

}
{
 
 \def\Pgamma      {\ensuremath{\gamma}\xspace}

 \def\Peta        {\ensuremath{\eta}\xspace}

 \def\Pmu         {\ensuremath{\mu}\xspace}

 \def\Ppi         {\ensuremath{\pi}\xspace}                 
                  
 \def\Prho        {\ensuremath{\rho}\xspace}

 \def\Pphi        {\ensuremath{\phi}\xspace}

 \def\Ppsi        {\ensuremath{\psi}\xspace}                 
 \def\Pomega      {\ensuremath{\omega}\xspace}                 
 \mathchardef\PDelta="7101
 \mathchardef\PXi="7104
 \mathchardef\PLambda="7103
 \mathchardef\PSigma="7106
 \mathchardef\POmega="710A
 \mathchardef\PUpsilon="7107
                  
 \def\PB      {\ensuremath{B}\xspace}                 
                  
 \def\PD      {\ensuremath{D}\xspace}

 \def\PJ      {\ensuremath{J}\xspace}                 
 \def\PK      {\ensuremath{K}\xspace}

 \def\Pb      {\ensuremath{b}\xspace}                 
 \def\Pc      {\ensuremath{c}\xspace}                 
                  
 \def\Pe      {\ensuremath{e}\xspace}

 \def\Pi      {\ensuremath{i}\xspace}

 \def\Pp      {\ensuremath{p}\xspace}

 \def\Ps      {\ensuremath{s}\xspace}

}

\makeatletter
\ifcase \@ptsize \relax
  \newcommand{\miniscule}{\@setfontsize\miniscule{4}{5}}
\or
  \newcommand{\miniscule}{\@setfontsize\miniscule{5}{6}}
\or
  \newcommand{\miniscule}{\@setfontsize\miniscule{5}{6}}
\fi
\makeatother

\DeclareRobustCommand{\optbar}[1]{\shortstack{{\miniscule (\rule[.5ex]{1.25em}{.18mm})}
  \\ [-.7ex] $#1$}}

\def\en         {{\ensuremath{\Pe^-}}\xspace}   
\def\ep         {{\ensuremath{\Pe^+}}\xspace}
 
\def\epem       {{\ensuremath{\Pe^+\Pe^-}}\xspace}

\def\mup        {{\ensuremath{\Pmu^+}}\xspace}
\def\mun        {{\ensuremath{\Pmu^-}}\xspace} 

\def\g      {{\ensuremath{\Pgamma}}\xspace}

\def\squark    {{\ensuremath{\Ps}}\xspace}

\def\cquark    {{\ensuremath{\Pc}}\xspace}

\def\bquark    {{\ensuremath{\Pb}}\xspace}

\def\pion   {{\ensuremath{\Ppi}}\xspace}
\def\piz    {{\ensuremath{\pion^0}}\xspace}

\def\kaon    {{\ensuremath{\PK}}\xspace}
  \def\Kbar    {{\kern 0.2em\overline{\kern -0.2em \PK}{}}\xspace}

\def\KorKbar    {\kern 0.18em\optbar{\kern -0.18em K}{}\xspace}

\def\KS      {{\ensuremath{\kaon^0_{\rm\scriptscriptstyle S}}}\xspace}

\def\Kstarz  {{\ensuremath{\kaon^{*0}}}\xspace}
\def\Kstarzb {{\ensuremath{\Kbar{}^{*0}}}\xspace}
\def\Kstar   {{\ensuremath{\kaon^*}}\xspace}

  \def\Dbar    {{\kern 0.2em\overline{\kern -0.2em \PD}{}}\xspace}
\def\D       {{\ensuremath{\PD}}\xspace}

\def\DorDbar    {\kern 0.18em\optbar{\kern -0.18em D}{}\xspace}

\def\Dm      {{\ensuremath{\D^-}}\xspace}

\def\B       {{\ensuremath{\PB}}\xspace}
\def\Bbar    {{\ensuremath{\kern 0.18em\overline{\kern -0.18em \PB}{}}}\xspace}

\def\BorBbar    {\kern 0.18em\optbar{\kern -0.18em B}{}\xspace}
\def\Bz      {{\ensuremath{\B^0}}\xspace}
\def\Bzb     {{\ensuremath{\Bbar{}^0}}\xspace}

\def\Bd      {{\ensuremath{\B^0}}\xspace}
\def\Bs      {{\ensuremath{\B^0_\squark}}\xspace}

\def\Bdb     {{\ensuremath{\Bbar{}^0}}\xspace}

\def\jpsi     {{\ensuremath{{\PJ\mskip -3mu/\mskip -2mu\Ppsi\mskip 2mu}}}\xspace}

  \def\Y#1S{\ensuremath{\PUpsilon{(#1S)}}\xspace}

\def\Lz          {{\ensuremath{\PLambda}}\xspace}
\def\Lbar        {{\ensuremath{\kern 0.1em\overline{\kern -0.1em\PLambda}}}\xspace}
\def\LorLbar    {\kern 0.18em\optbar{\kern -0.18em \PLambda}{}\xspace}

\def\Lb      {{\ensuremath{\Lz^0_\bquark}}\xspace}

\newcommand{\decay}[2]{\ensuremath{#1\!\to #2}\xspace}         
\def\ra                 {\ensuremath{\rightarrow}\xspace}
\def\to                 {\ensuremath{\rightarrow}\xspace}

\def\qsq       {{\ensuremath{q^2}}\xspace}

\def\CP                {{\ensuremath{C\!P}}\xspace}

\def\BdToKstmm    {\decay{\Bd}{\Kstarz\mup\mun}}
\def\BdToKstll    {\decay{\B}{\Kstar \ell^+\ell^-}}

\def\BdKstGam     {\decay{\Bd}{\Kstarz \g}}

\def\BdKstee  {\decay{\Bd}{\Kstarz\epem}}

\def\AFB      {\ensuremath{A_{\mathrm{FB}}}\xspace}
\def\AT#1     {\ensuremath{A_{\mathrm{T}}^{#1}}\xspace}           

\def\C#1      {\ensuremath{\mathcal{C}_{#1}}\xspace}                       
\def\Cp#1     {\ensuremath{\mathcal{C}_{#1}^{'}}\xspace}                    
\def\Ceff#1   {\ensuremath{\mathcal{C}_{#1}^{\mathrm{(eff)}}}\xspace}        
\def\Cpeff#1  {\ensuremath{\mathcal{C}_{#1}^{'\mathrm{(eff)}}}\xspace}       
\def\Ope#1    {\ensuremath{\mathcal{O}_{#1}}\xspace}                       
\def\Opep#1   {\ensuremath{\mathcal{O}_{#1}^{'}}\xspace}                    

\newcommand{\tev}{\ifthenelse{\boolean{inbibliography}}{\ensuremath{~T\kern -0.05em eV}\xspace}{\ensuremath{\mathrm{\,Te\kern -0.1em V}}}\xspace}
\newcommand{\gev}{\ensuremath{\mathrm{\,Ge\kern -0.1em V}}\xspace}
\newcommand{\mev}{\ensuremath{\mathrm{\,Me\kern -0.1em V}}\xspace}
\newcommand{\kev}{\ensuremath{\mathrm{\,ke\kern -0.1em V}}\xspace}
\newcommand{\ev}{\ensuremath{\mathrm{\,e\kern -0.1em V}}\xspace}
\newcommand{\gevc}{\ensuremath{{\mathrm{\,Ge\kern -0.1em V\!/}c}}\xspace}
\newcommand{\mevc}{\ensuremath{{\mathrm{\,Me\kern -0.1em V\!/}c}}\xspace}
\newcommand{\gevcc}{\ensuremath{{\mathrm{\,Ge\kern -0.1em V\!/}c^2}}\xspace}
\newcommand{\mevcc}{\ensuremath{{\mathrm{\,Me\kern -0.1em V\!/}c^2}}\xspace}

\def\mm   {\ensuremath{\rm \,mm}\xspace}

\def\mum  {\ensuremath{{\,\upmu\rm m}}\xspace}

\def\invfb   {\ensuremath{\mbox{\,fb}^{-1}}\xspace}

\newcommand{\chisq}{\ensuremath{\chi^2}\xspace}

\def\deriv {\ensuremath{\mathrm{d}}}

\def\gsim{{~\raise.15em\hbox{$>$}\kern-.85em
          \lower.35em\hbox{$\sim$}~}\xspace}
\def\lsim{{~\raise.15em\hbox{$<$}\kern-.85em
          \lower.35em\hbox{$\sim$}~}\xspace}

\newcommand{\Real}{\ensuremath{\mathcal{R}e}\xspace}
\newcommand{\Imag}{\ensuremath{\mathcal{I}m}\xspace}

\def\sPlot{\mbox{\em sPlot}\xspace}

\def\ptot       {\mbox{$p$}\xspace}
\def\pt         {\mbox{$p_{\rm T}$}\xspace}

\def\evtgen     {\mbox{\textsc{EvtGen}}\xspace}

\def\geant      {\mbox{\textsc{Geant4}}\xspace}

\def\photos     {\mbox{\textsc{Photos}}\xspace}

\def\pythia     {\mbox{\textsc{Pythia}}\xspace}

\def\tell1  {TELL1\xspace}
\def\ukl1   {UKL1\xspace}

\newcommand{\ie}{\mbox{\itshape i.e.}\xspace}

%% file: title-LHCb-PAPER.tex

\begin{titlepage}
\pagenumbering{roman}

\vspace*{-1.5cm}
\centerline{\large EUROPEAN ORGANIZATION FOR NUCLEAR RESEARCH (CERN)}
\vspace*{1.5cm}
\hspace*{-0.5cm}
\begin{tabular*}{\linewidth}{lc@{\extracolsep{\fill}}r}
\ifthenelse{\boolean{pdflatex}}
{\vspace*{-2.7cm}\mbox{\!\!\!\includegraphics[width=.14\textwidth]{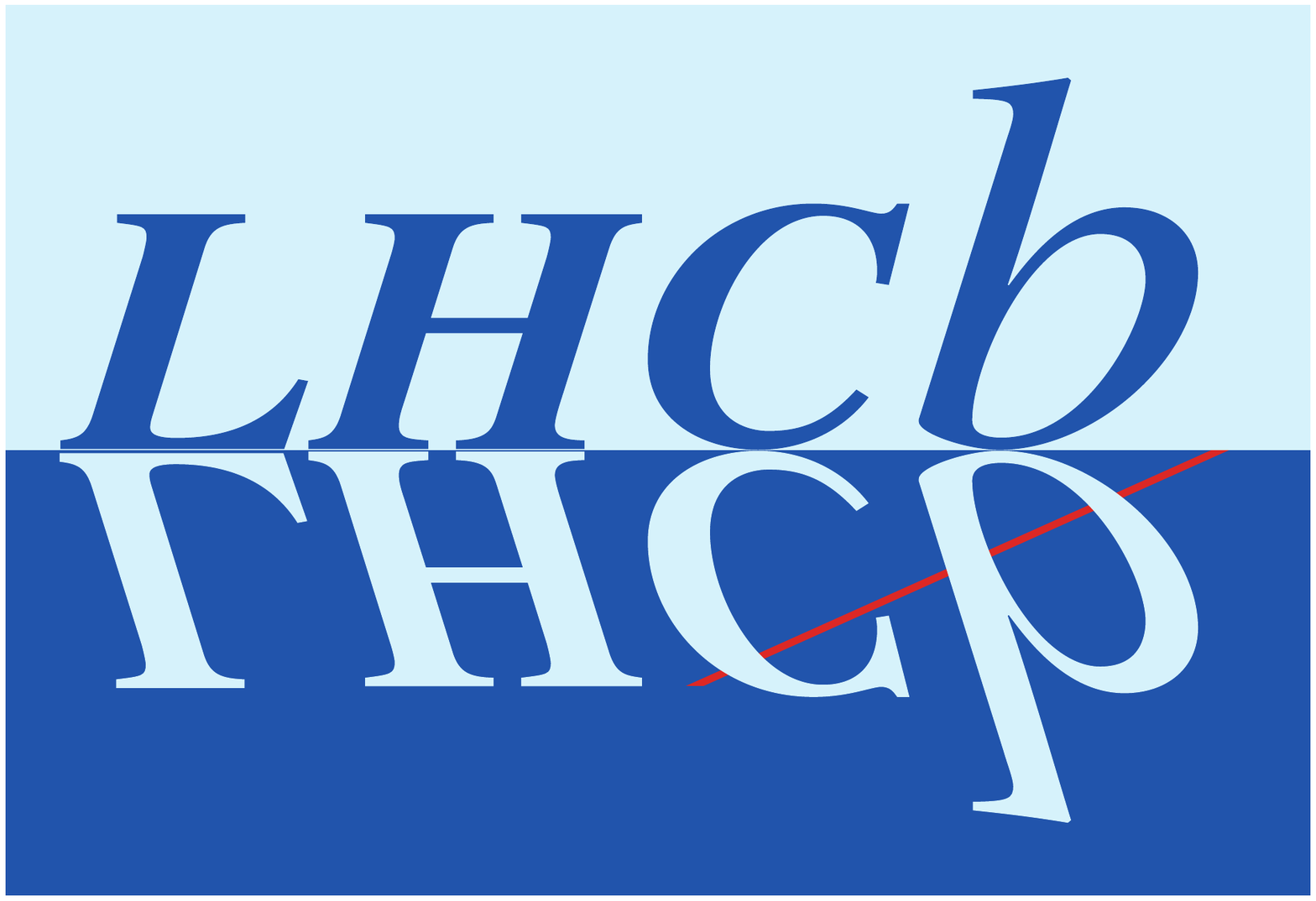}} & &}%
{\vspace*{-1.2cm}\mbox{\!\!\!\includegraphics[width=.12\textwidth]{lhcb-logo.eps}} & &}%
\\
 & & CERN-PH-EP-2014-301 \\  
 & & LHCb-PAPER-2014-066 \\  
 & & 3 March 2015 \\ 
 & & \\
\end{tabular*}

\vspace*{2.0cm}

{\bf\boldmath\huge
\begin{center}
  Angular analysis of the $B^0 \rightarrow K^{*0} e^+ e^-$ decay in the low-$q^2$ region
\end{center}
}

\vspace*{1.0cm}

\begin{center}
The LHCb collaboration\footnote{Authors are listed at the end of this paper.}
\end{center}

\vspace{\fill}

\begin{abstract}
  \noindent
  An angular analysis of the $B^0 \rightarrow K^{*0} e^+ e^-$ decay is performed using a data sample,
  corresponding to an integrated luminosity of 3.0 ${\mbox{fb}^{-1}}$, collected by the LHCb experiment in $pp$ collisions at centre-of-mass energies of 7 and 8 TeV during 2011 and 2012.
  For the first time several observables are measured in the dielectron mass squared ($q^2$) interval between 0.002 and 1.120${\mathrm{\,Ge\kern -0.1em V^2\!/}c^4}$. The 
  angular observables $F_{\mathrm{L}}$ and $A_{\mathrm{T}}^{\mathrm{Re}}$ which are related to the $K^{*0}$ polarisation and to the lepton forward-backward asymmetry, are measured to be $F_{\mathrm{L}}= 0.16 \pm 0.06 \pm0.03$ and $A_{\mathrm{T}}^{\mathrm{Re}} = 0.10 \pm 0.18 \pm 0.05$, where the first uncertainty is statistical and the second systematic. The angular
  observables $A_{\mathrm{T}}^{(2)}$ and $A_{\mathrm{T}}^{\mathrm{Im}}$ which are sensitive to the photon polarisation in this $q^2$ range, are found to be $A_{\mathrm{T}}^{(2)} = -0.23 \pm 0.23 \pm 0.05$ and $A_{\mathrm{T}}^{\mathrm{Im}} =0.14 \pm 0.22 \pm 0.05$. The results are consistent with Standard Model predictions. 
\end{abstract}

\vspace*{1.0cm}

\begin{center}
JHEP  {\bf 04} (2015) 064.  
\end{center}

\vspace{\fill}

{\footnotesize 
\centerline{\copyright~CERN on behalf of the \lhcb collaboration, license \href{http://creativecommons.org/licenses/by/4.0/}{CC-BY-4.0}.}}
\vspace*{2mm}

\end{titlepage}


\newpage
\setcounter{page}{2}
\mbox{~}
\cleardoublepage

%% file: Introduction.tex
\section{Introduction}

\begin{figure}[b]
  \centering
  \includegraphics[angle=0,width=0.48\textwidth]{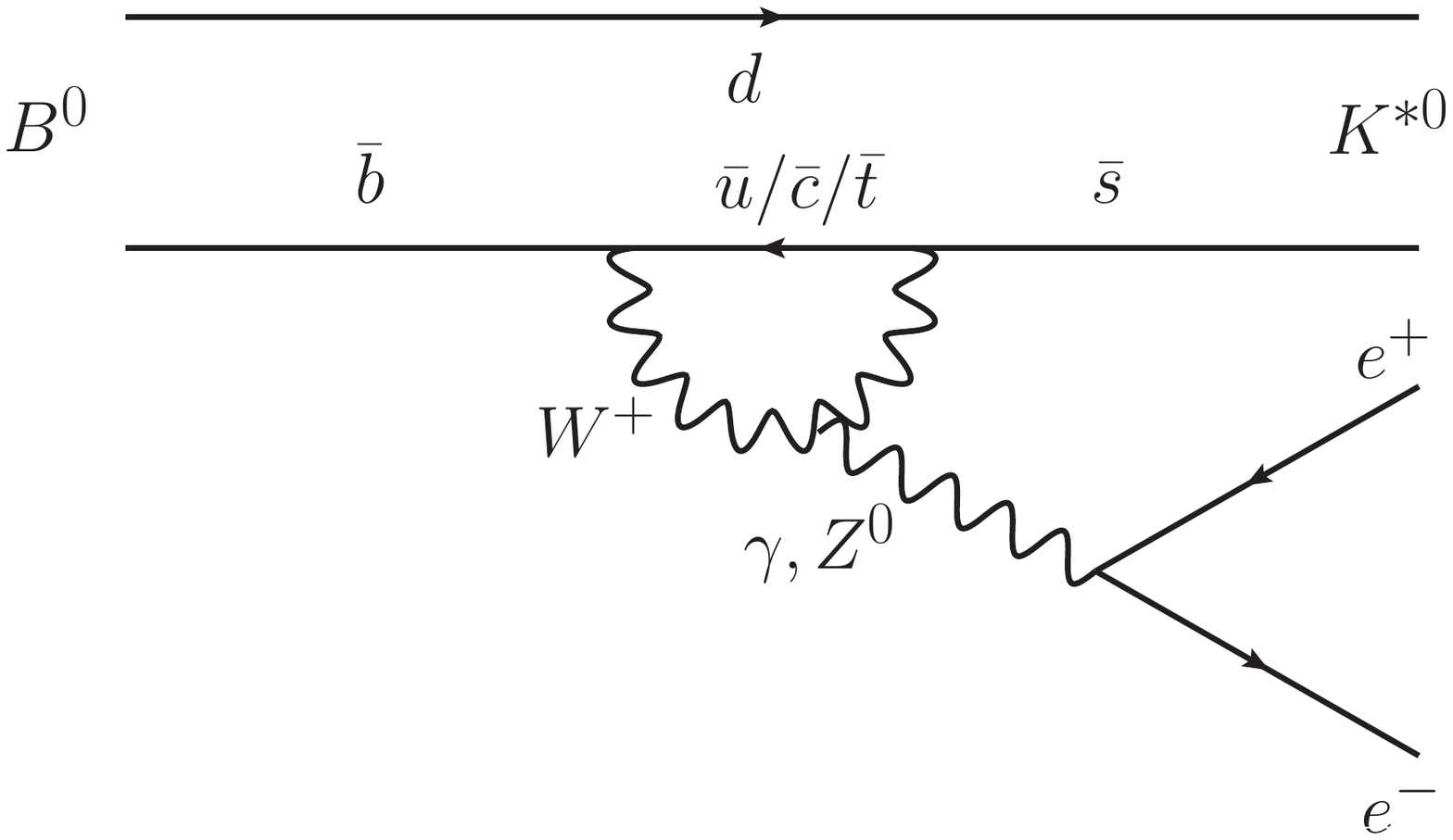}
  \includegraphics[angle=0,width=0.48\textwidth]{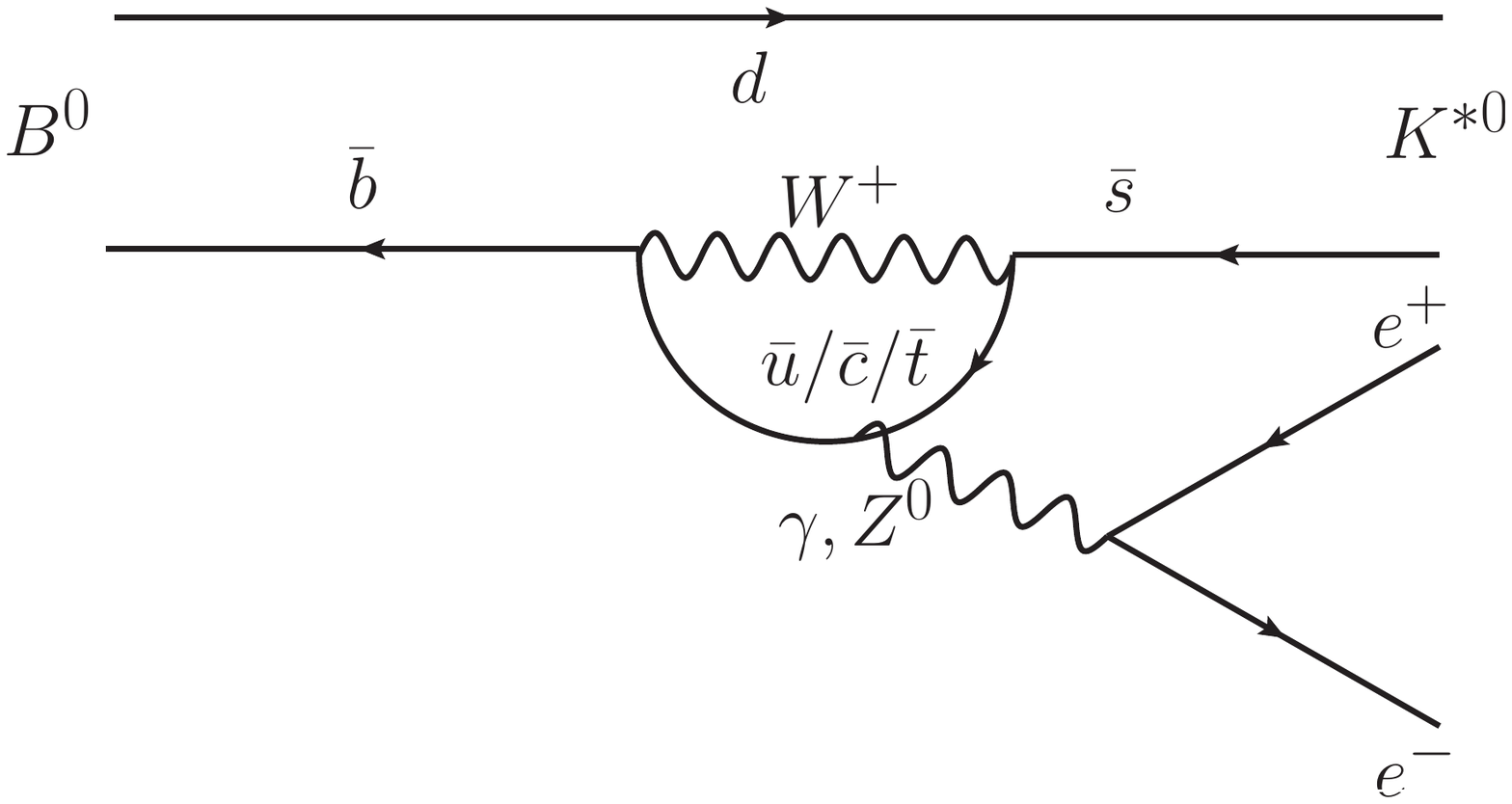}
  \vskip -6cm
  \includegraphics[angle=0,width=0.48\textwidth]{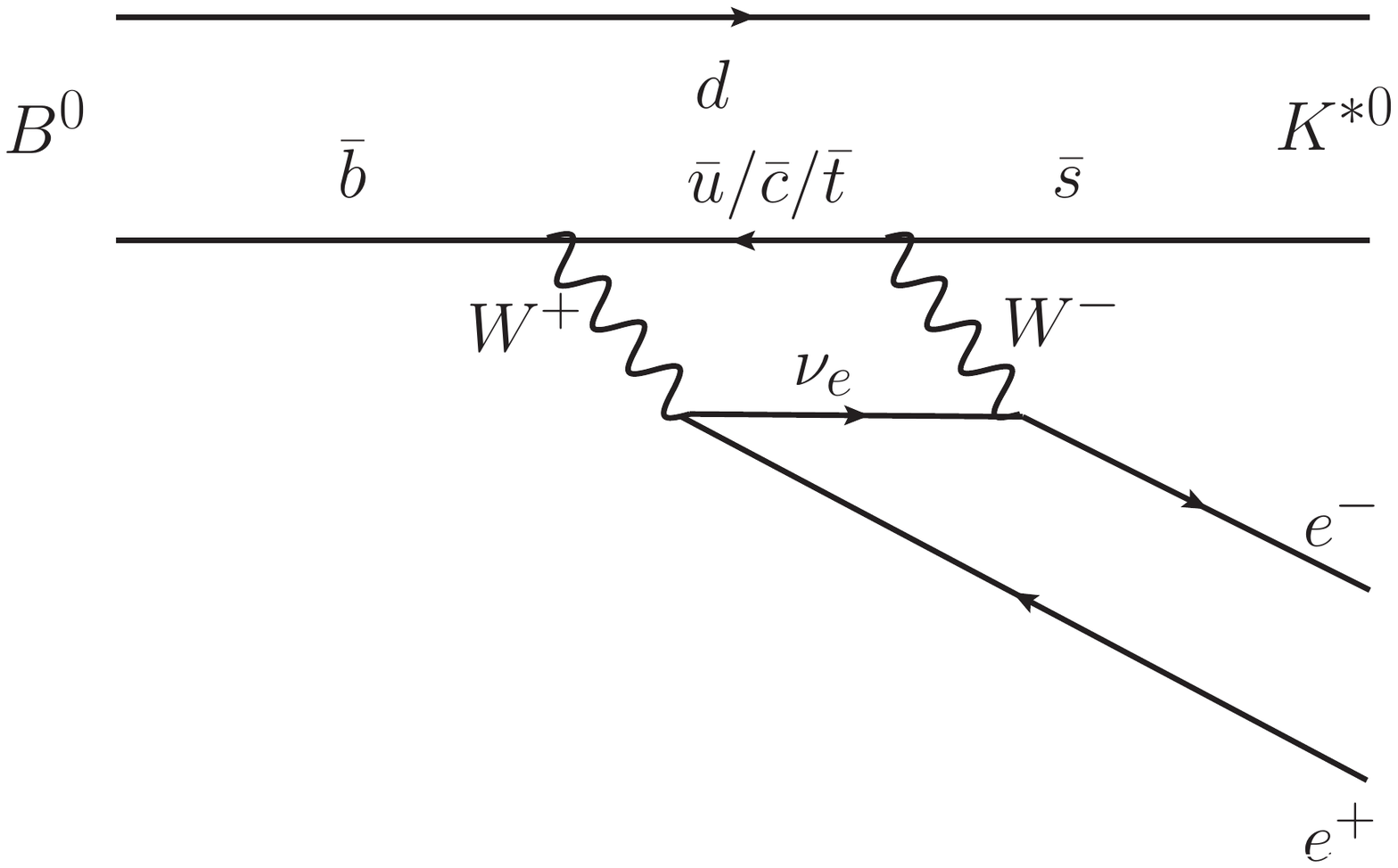}
   \vskip -6cm
  \caption{Dominant Standard Model Feynman graphs for the electroweak loop and box diagrams involved in the \BdToeeKst decay.}
  \label{fig:feynman}
\end{figure}

The \BdToeeKst decay is a flavour changing neutral current process that is mediated by electroweak box and loop diagrams in the Standard Model (SM). Charge conjugation is implied throughout this paper unless stated otherwise and the \Kstarz represents the $\kaon^{*0}(892)$, reconstructed as \KstToKPi. The angular distribution of the $K^+ \pi^- \epem$ system is particularly sensitive to contributions from non-SM physics (NP). The leading SM diagrams are shown in Fig.~\ref{fig:feynman};
the relative contribution of each of the diagrams varies with the dilepton invariant mass. In the region where the dilepton invariant mass squared~(\qsq) is less than 6\gevgevcccc, some theoretical uncertainties from long distance contributions are greatly reduced, thereby allowing more control over the SM prediction and increasing sensitivity to any NP effect~\cite{Kruger:2005ep,Becirevic:2011bp}. Furthermore, the contribution from a virtual photon coupling to the lepton pair dominates in the very low \qsq region, allowing measurement of the helicity of the photon in  \decay{b}{s\g} transitions~\cite{Grossman:2000rk,Jager:2012uw}.  In the SM, this photon is predominantly left-handed, with a
small right-handed component arising from the mass of the $s$ quark and long distance effects. In contrast, in many extension of the SM, NP may manifest as a large right handed current, see for example Refs.~\cite{Everett:2001yy, Foster:2006ze, Lunghi:2006hc,Goto:2007ee}. 

The \qsq region below 1\gevgevcccc has previously been studied through the analysis of the \BdToKstll($\ell = \Pe, \mu$)~\cite{Wei:2009zv,Aaltonen:2011ja,LHCb-PAPER-2013-019}. Experimentally, an analysis with muons rather than electrons in the final state produces a much higher yield at \lhcb. This is 
primarily due to the distinctive signature that muons provide, which is efficiently exploited in the online selection, together with the 
better mass and energy resolutions and higher reconstruction efficiency of dimuon decays. However, as outlined in Ref.~\cite{Lefrancois:1179865}, dielectron decays at low \qsq provide greater sensitivity
to the photon polarisation and therefore to the \C7 and \Cp{7} Wilson coefficients, which are associated with the left-handed and right-handed electromagnetic operators, respectively~\cite{Grossman:2000rk}. Due to the muon mass, the virtual photon contribution in dimuon decays is suppressed relative to dielectron decays. Additionally, the formalism for the \BdToeeKst decay is greatly simplified as the electron mass can be neglected. Indeed, the decay with electrons allows for an angular analysis down to a \qsq of 0.0004\gevgevcccc. However, above a \qsq of 1\gevgevcccc, 
the muon mass terms become negligible and the electron and muon modes have essentially the same functional dependence on the Wilson coefficients (within the lepton flavour universality assumption).   

This work is based on a previous analysis performed by the \lhcb collaboration to measure the \BdToeeKst branching fraction with an integrated luminosity of 1.0\invfb ~\cite{LHCb-PAPER-2013-005}, with the selection re-optimised for the angular analysis. 

The partial decay width of the \BdToeeKst decay can be described in terms of \qsq and three angles, $\theta_{\ell},\theta_{K}$ and $\phi$.  The angle $\theta_{\ell}$ is defined as the angle between the direction of the \ep (\en) and the direction opposite to that of the \Bz (\Bzb) meson in the dielectron rest frame. The angle $\theta_{K}$ is defined as the angle between the direction of the kaon and the direction opposite to that of the \Bz (\Bzb) meson in the \Kstarz (\Kstarzb) rest frame. The angle $\phi$ is the angle between the plane containing the \ep and \en and the plane containing the kaon and pion from the \Kstarz (\Kstarzb) in the \Bz (\Bzb) rest frame.  The basis is designed such that the angular definition for the \Bzb decay is a \CP transformation of that for the \Bz decay. These definitions are identical to those used for the \BdToKstmm analysis~\cite{LHCb-PAPER-2013-019}.
As in Ref.~\cite{LHCb-PAPER-2013-019}, the angle $\phi$ is transformed such that $\phit=\phi+\pi$ if $\phi<0$, to compensate for the limited signal yield. This transformation cancels out the terms that have a $\sin\phi$ or $\cos\phi$ dependence and simplifies the angular expression without any loss of sensitivity to the remaining observables.
In the limit of massless leptons and neglecting the $K^+ \pi^-$ S-wave contribution, which is expected to be negligible\footnote{Using Refs~\cite{Kruger:2005ep,Lu:2011jm}  it can be shown that the ratio of the S-wave fraction to the fraction of longitudinal polarisation of the \Kstarz is constant as function of \qsq in the 0-6  \gevgevcccc range. } at low \qsq with the current sample size~\cite{Lu:2011jm}, the \BdToeeKst angular distribution reads as
\begin{equation}
  \label{eq:AngDistr}
  \begin{split}
  \left. \frac{1}{\deriv(\Gamma+\bar\Gamma)/\deriv q^2} \frac{\deriv^4(\Gamma+\bar\Gamma)}{\deriv q^2\,\deriv{\cos\theta_\ell}\,\deriv{\ctk}\,\deriv{\phit}} = 
\frac{9}{16\pi} \right. & \left[ \frac{}{} \frac{3}{4}(1-\FL)\sin^2\theta_K+\FL\cos^2\theta_K ~+  \right.\\
  &\left. ~\frac{}{} \left( \frac{1}{4}(1-\FL)\sin^2\theta_K - \FL\cos^2\theta_K \right)\cos2\theta_\ell \right. ~+\\
  &\left. ~\frac{}{} \frac{1}{2}(1-\FL)\ATD\sin^2\theta_K \sin^2\theta_\ell\cos 2\phit  \right. ~+ \\
  & \left. ~\frac{}{} (1-\FL)\ATRe\sin^2\theta_K \ctl \right. ~+ \\
  & \left. ~\frac{}{} \frac{1}{2}(1-\FL) \ATIm\sin^2\theta_K \sin^2\theta_\ell\sin 2\phit  ~\right]. 
\end{split}
\end{equation}
The four angular observables \FL, \ATD, \ATRe and \ATIm are related to the transversity amplitudes through~\cite{Becirevic:2011bp}
\begin{equation}
  \label{eq:physPars}
  \begin{split}
  \FL &=\frac{|A_0|^2}{|A_0|^2+|A_{||}|^2 + |A_\perp|^2}\\ 
  \ATD &= \frac{|A_\perp|^2-|A_{||}|^2}{|A_\perp|^2+|A_{||}|^2}\\ 
  \ATRe &= \frac{2\Real(A_{||L}A^*_{\perp L} + A_{||R}A^*_{\perp R})}{|A_{||}|^2 + |A_\perp|^2}\\
  \ATIm &= \frac{2\Imag(A_{||L}A^*_{\perp L} + A_{||R}A^*_{\perp R})}{|A_{||}|^2 + |A_\perp|^2}, 
\end{split}
\end{equation}
where $|A_0|^2=|A_{0L}|^2+|A_{0R}|^2$, 
$|A_\perp |^2=|A_{\perp L}|^2+|A_{\perp R}|^2$ and 
$|A_{||} |^2=|A_{|| L}|^2+|A_{|| R}|^2$. The amplitudes $A_0$, $A_{||}$ and $A_\perp$ correspond to different polarisation states of the \Kstarz in the decay. The labels $L$ and $R$ refer to the left and right chirality of the dielectron system.

Given the definition of \phit, the observable \ATD is averaged between \Bz and \Bzb decays, while \ATIm corresponds to a \CP asymmetry~\cite{Bobeth:2008ij}.
The observable \FL is the longitudinal polarisation of the \Kstarz and is expected to be small at low \qsq, since the virtual photon is then quasi-real and therefore transversely polarised. The observable \ATRe is related to the forward-backward asymmetry \AFB by $\ATRe = \frac{4}{3}\AFB / (1-\FL)$~\cite{Becirevic:2011bp}. 
The observables \ATD and \ATIm, in the limit $\qsq \to 0$, can be expressed as simple functions of the \C7 and \Cp{7} coefficients~\cite{Becirevic:2011bp}
\begin{equation}
  \label{eq:qSqToZero} 
\ATD (\qsq \to 0) = \frac{2 \Real (\C7 \mathcal{C}_7^{' \ast}) } {| \C7 |^2 + |\mathcal{C}_7^{'}|^2 }  \ \ \mathrm {and} \  \  \ATIm (\qsq \to 0) = \frac{2 \Imag (\C7 \mathcal{C}_7^{' \ast}) } {| \C7 |^2 + |\mathcal{C}_7^{'}|^2 }.
 \end{equation}
These measurements therefore provide information on photon polarisation amplitudes, similar to that obtained by the \CP asymmetry measured through time-dependent analyses in \BdToKstPizGamma decays~\cite{Aubert:2008gy,Ushiroda:2006fi}. 

This paper presents measurements of \FL, \ATD, \ATIm and \ATRe of the \BdKstee decay in the bin corresponding to a reconstructed \qsq from 0.0004 to 1\gevgevcccc.

%% file: Detector.tex
\section{The \lhcb detector and data set}
\label{Sec:Detector}
The study reported here is based on $pp$ collision data, corresponding to an integrated luminosity of 3.0\invfb, collected at the Large Hadron Collider 
(LHC) with the \lhcb detector~\cite{Alves:2008zz,LHCb-DP-2014-002} at centre-of-mass energies of 7 and 8 \tev during 2011 and 2012.  The \lhcb 
detector is a single-arm forward
spectrometer covering the \mbox{pseudorapidity} range $2<\eta <5$,
designed for the study of particles containing \bquark or \cquark
quarks. The detector includes a high-precision tracking system
consisting of a silicon-strip vertex detector surrounding the $pp$
interaction region~\cite{LHCb-DP-2014-001}, a large-area silicon-strip detector located
upstream of a dipole magnet with a bending power of about
$4{\rm\,Tm}$, and three stations of silicon-strip detectors and straw
drift tubes~\cite{LHCb-DP-2013-003} placed downstream of the magnet.
The tracking system provides a measurement of momentum, \ptot,  with
a relative uncertainty that varies from 0.5\% at low momentum to 1.0\% at 200\gevc.
The minimum distance of a track to a primary vertex, the impact parameter (IP), is measured with a resolution of $(15+29/\pt)\mum$,
where \pt is the component of the momentum transverse to the beam, in \gevc.
Different types of charged hadrons are distinguished using information
from two ring-imaging Cherenkov detectors~\cite{LHCb-DP-2012-003}. Photons, electrons and
hadrons are identified by a calorimeter system consisting of
scintillating-pad and preshower detectors, an electromagnetic
calorimeter (ECAL) and a hadronic calorimeter. Muons are identified by a
system composed of alternating layers of iron and multiwire
proportional chambers~\cite{LHCb-DP-2012-002}.

The trigger~\cite{LHCb-DP-2012-004} consists of a
hardware stage, based on information from the calorimeter and muon
systems, followed by a software stage, which applies a full event
reconstruction.
For signal candidates to be considered in this analysis, all tracks from the \BdKstee decay must have hits in the vertex detector and at least one of the tracks from the \BdKstee decay must
meet the requirements of the hardware electron or hadron triggers, or the hardware trigger must be fulfilled independently of any of the decay products
of the signal \Bz candidate (usually triggering on the other \bquark hadron in the event). The hardware electron trigger requires the presence of an ECAL 
cluster with a minimum transverse energy between 2.5\gev and 2.96\gev depending on the data taking period. The hardware hadron trigger requires the presence of a 
cluster in the hadron calorimeter with a transverse energy greater than 3.5\gev. 
The software trigger requires a two-, three- or four-track
  secondary vertex with a significant displacement from the primary
  $pp$ interaction vertices~(PVs). At least one charged particle
  must have a transverse momentum $\pt > 1.7\gevc$ and be
  inconsistent with originating from the PV.
  A multivariate algorithm~\cite{BBDT} is used for
  the identification of secondary vertices consistent with the decay
  of a \bquark hadron.
  
 Samples of simulated \BdToeeKst events are used to determine the efficiency to trigger, reconstruct and select signal events. 
 In addition, specific samples of simulated events are utilised to estimate the contribution from exclusive backgrounds and to model their mass and angular distributions. 
 The $pp$ collisions are generated using
\pythia~\cite{Sjostrand:2006za,*Sjostrand:2007gs}
with a specific \lhcb
configuration~\cite{LHCb-PROC-2010-056}.  Decays of hadronic particles
are described by \evtgen~\cite{Lange:2001uf}, in which final-state
radiation is generated using \photos~\cite{Golonka:2005pn}. The
interaction of the generated particles with the detector, and its
response, are implemented using the \geant
toolkit~\cite{Allison:2006ve, *Agostinelli:2002hh} as described in
Ref.~\cite{LHCb-PROC-2011-006}.
The simulated samples are corrected
for known differences between data and simulation in particle identification~\cite{LHCb-DP-2012-003}, detector occupancy and hardware trigger efficiency.

%% file: Selection.tex
\section{Selection of signal candidates}
\label{Sec:Selection}
Bremsstrahlung radiation, if not accounted for,  would worsen the \Bz mass resolution. If the radiation occurs downstream of the dipole magnet, the
momentum of the electron is correctly measured and the photon energy is deposited in the
same calorimeter cell as the electron. If
photons are emitted upstream of the magnet, the electron momentum is evaluated after
photon emission, and the measured \Bz mass is shifted. In general, these bremsstrahlung photons deposit their energy in 
different calorimeter cells than those hit by the
electron.  In both cases, the ratio of the energy detected in the ECAL to the momentum measured by the tracking system, an important variable in identifying electrons, remains unbiased. 
To improve the momentum reconstruction, a dedicated bremsstrahlung recovery is used. Contributions from photon candidates, neutral clusters with transverse energy greater than 75\mev, found within a region of the ECAL defined by the extrapolation of the electron track upstream of the magnet, are added to the measured electron momentum. 

\begin{figure}[tbp]
  \begin{center}
  \includegraphics[width=0.6\textwidth]{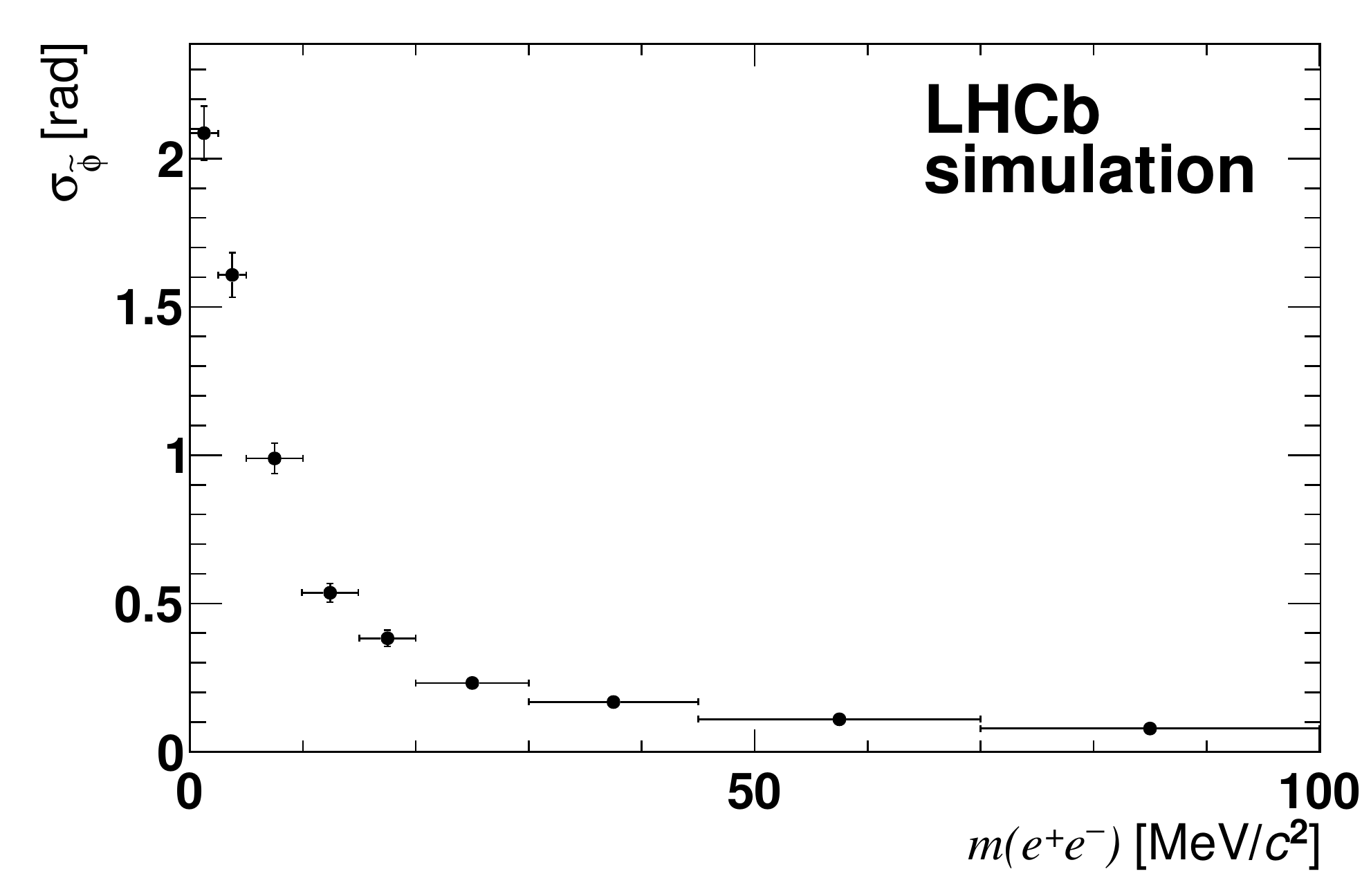}
 \vspace*{-1.0cm}
  \end{center}
  \caption{\small Resolution on the \phit angle as a function of the \epem invariant mass as obtained from \lhcb simulated events.}
  \label{fig:ResoPhi}
\end{figure}

Oppositely charged electron pairs formed from tracks with \pt exceeding 350\mevc and with a good-quality vertex are used to form signal candidates. If the same bremsstrahlung photon is associated with both the $\Pe^+$ and the $\Pe^-$, its energy is added randomly to one of the tracks. The reconstructed \epem invariant mass is required to be in the range 20--1000 \mevcc ($0.0004<\qsq<1\gevgevcccc$).  The choice of the lower bound is a compromise between the gain in sensitivity to the photon polarisation from measuring as low as possible in \qsq\ and a degradation of the resolution in $\phit$ as \qsq decreases, due to multiple scattering, as shown in Fig.~\ref{fig:ResoPhi}. The lower bound requirement at 20\mevcc on the \epem invariant mass also serves to reduce the background from \BdKstGam decays followed by a photon conversion in the material, noted below as \BdKstGamToee. 

Candidate \Kstarz mesons are reconstructed in the $\Kstarz \to K^+ \pi^-$ mode where the \pt of the $K^+$ ($\pi^-$) meson is required to be larger than 400 (300)\mevc 
and charged pions and kaons are identified using information from the RICH detectors. 

Candidate \Kstarz mesons and \epem pairs are required to have a common good-quality vertex to form \Bd candidates. 
When more than one PV is reconstructed, the one giving the smallest IP \chisq for the \Bd candidate is chosen. 
The  reconstructed decay vertex of the \Bd candidate is required to be significantly separated from the PV and the candidate momentum direction to be consistent with its direction of flight from the PV. 
The \Bd mass resolution, the angular acceptance and the rates of physics and combinatorial backgrounds depend on how the event was triggered. The data sample is therefore divided into three mutually exclusive categories: events for which one of the electrons from the \Bz decay satisfies the hardware electron trigger, 
events for which one of the hadrons from the \Bz decay satisfies the hardware hadron trigger and events triggered by activity in the event not due to any of the signal decay particles. 

In order to maximise the signal efficiency while reducing the high level of combinatorial background, a multivariate classifier based on a boosted decision tree algorithm \mbox{(BDT)~\cite{Breiman,AdaBoost}} is used.  The signal training sample is composed of simulated \BdKstee events and the background training sample is taken from the upper invariant mass 
sideband ($\meeKstar > 5600\mevcc$) of \BdKstee decays reconstructed in half of the data sample. Two separate BDTs are used, one each for half of the data sample. They are optimised separately and applied to the complementary half of the data in order to avoid any potential bias due to the use of the data upper sideband for the background sample. The BDT uses information about the event kinematic properties, vertex and track quality, IP and \pt of the tracks, flight distance from the PV as well as information about 
isolation of the final state particles.\footnote{The isolation is defined as the number of good two-track vertices that one of the candidate signal tracks can make with any other track in the event~\cite{Abulencia:2005pw}.}
The selection is optimised to maximise $N_S/\sqrt{N_S+N_B}$ separately for the three trigger categories and the two BDTs through a grid search of the set of criteria for the particle identification of the four final state particles and the BDT response. 
The background yield ($N_B$) is extrapolated into the signal range using the \meeKstar\ distribution outside a $\pm 300 \mevcc$ window around the known \Bd mass. The expected signal yield ($N_S$) is obtained 
using the \BdKstee simulation and the known \BdKstee branching fraction~\cite{LHCb-PAPER-2013-005}, and correcting for data-to-simulation differences in the selection efficiency obtained using the well known \BdToJPsieeKst decay. 
 The efficiency of this requirement on the selected signal is 93\% while the background is reduced by two orders of magnitude. 
 The expected values for $N_S/\sqrt{N_S+N_B}$ range from 3.9 to 7.5 depending on the trigger category.

%% file: Backgrounds.tex
\section{Exclusive and partially reconstructed backgrounds}
\label{sec:Backgrounds}
Several sources of background are studied using samples of simulated events, corrected to reflect the difference in particle identification
performances between data and simulation. 

A large non-peaking background comes from the \decay{\Bd}{\Dm\ep\nu} decay, with  ${\Dm \ra \en { \overline \nu} \Kstarz}$ which has a combined
branching fraction about four orders of magnitude larger than that of the signal. 
 In the rare case where both neutrinos have low energies, the signal selection is ineffective at rejecting this
background which tends to peak towards $\ctl \approx 1$. In order to avoid any potential bias in the measurement 
of the \ATRe parameter, a symmetric requirement of $|\ctl|<0.8$ is applied to suppress this background, resulting in a loss of signal of the order of 10\%. 

To suppress  background from \BsToPhiee decays, with \decay{\phi}{K^+K^-}, where one of the kaons is misidentified as
a pion, the two-hadron invariant mass computed under the $K^+K^-$ hypothesis is required to be larger than 1040\mevcc.  

Background from the decay \LbToeeL is suppressed by rejecting events where the pion is consistent with being a proton, according to the information from the RICH detectors. 

The probability for a decay \decay{\Bd}{\Kstarz\epem} to be misidentified as \decay{\Bdb}{\Kstarzb\epem} is estimated to be 1.1 \% using simulated events and this background is therefore neglected. 

Another important source of background comes from the \BdKstGam decay, where the photon converts into an \epem pair. In \lhcb, approximately 40\% of the photons convert before reaching the calorimeter, and although only about 10\% are reconstructed as an \epem pair with hits in the vertex detector, the resulting mass of the \Bd candidate peaks in the signal region. 
Two very effective criteria for suppressing this background are the minimum requirement on the \epem invariant mass, $m(\epem) > 20\mevcc$, and a requirement that the uncertainty of the reconstructed $z$ coordinate of the \epem pair, $\sigma_z(\epem)$, is less than  $30\mm$. These requirements reject more than $99\%$ of simulated \BdKstGam events. The remaining contamination is estimated by normalising the simulated \BdKstGamToee to the observed yield without the $\sigma_z(\epem)$ criterion and requiring the \epem invariant mass to be lower than 5~\mevcc . 
The residual contamination from \BdKstGam decays is $(3.8 \pm 1.9)\%$ of the signal yield. Part of this background comes from low-mass \epem pairs that are reconstructed at larger masses due to multiple scattering. The remainder comes from direct Bethe-Heitler pair-production at masses larger than 20\mevcc. To obtain an accurate estimate of this component, the \geant simulation is reweighted as a function of the true \epem mass to match the distribution of Ref.~\cite{Borsellino:1953zz} 
since \geant does not model correctly the high-mass \epem pair production. 

 Another possible source of contamination is the decay \BdKstV where $V$ is a \Prho, \Pomega or \Pphi meson. Expected rates for these backgrounds have been evaluated in 
 Refs.~\cite{Korchin:2010uc, Jager:2012uw}. The effects of direct decays or interference with the signal decay are found to be negligible after integrating over the \qsq range.

 Partially reconstructed (PR) backgrounds arising from $\decay{\Bd}{\Kstarz\epem X}$ decays, where one or more of the decay products ($X$) from the \Bd decay is not reconstructed, are also taken into account. 
These incomplete events are mostly due to decays involving higher \Kstar resonances, hereafter referred to as $\kaon^{**}$. The decays \BdKstEta and \BdKstPiz are also studied and several cases  are considered: the case when the \epem pair comes from a converted photon in the material, the case when the $\Pe^+$ and $\Pe^-$ originate from the conversions of the two photons and finally the case of the Dalitz decay of the \Peta or the \piz. They contribute about 25\% of the PR background in the angular fitting domain.

%% file: MassFit.tex
\section{Fit to the $K^+\pi^- \epem$ invariant mass distribution \label{sec:MassFit}}
In a first step, a mass fit over a wide mass range, from 4300 to 6300\mevcc, is performed to estimate the size of the \BdToeeKst signal, the combinatorial background and the PR background. 
The fractions of each component are determined from unbinned maximum likelihood fits to the mass distributions separately for each trigger category. The mass distribution of each category is fitted to a sum of probability density functions (PDFs), modelling the different components. Following the strategy of Ref.~\cite{LHCb-PAPER-2014-024}, the signal PDF depends on the number of neutral clusters that
are added to the dielectron candidate to correct for the effects of bremsstrahlung. The signal is described by the sum of a Crystal Ball function~\cite{Skwarnicki:1986xj} (CB) and a wide Gaussian function accounting for the cases where background photons have been associated; the CB function accounts for over 90\% of the total signal PDF. 
The shape of the combinatorial background is parameterised by an exponential function. Finally, the shape of the PR background is described by non-parametric PDFs~\cite{Cranmer:2000du} determined from fully simulated events passing the selection. 

\begin{figure}[tbp]
  \centering
  \includegraphics[angle=0,width=0.495\textwidth]{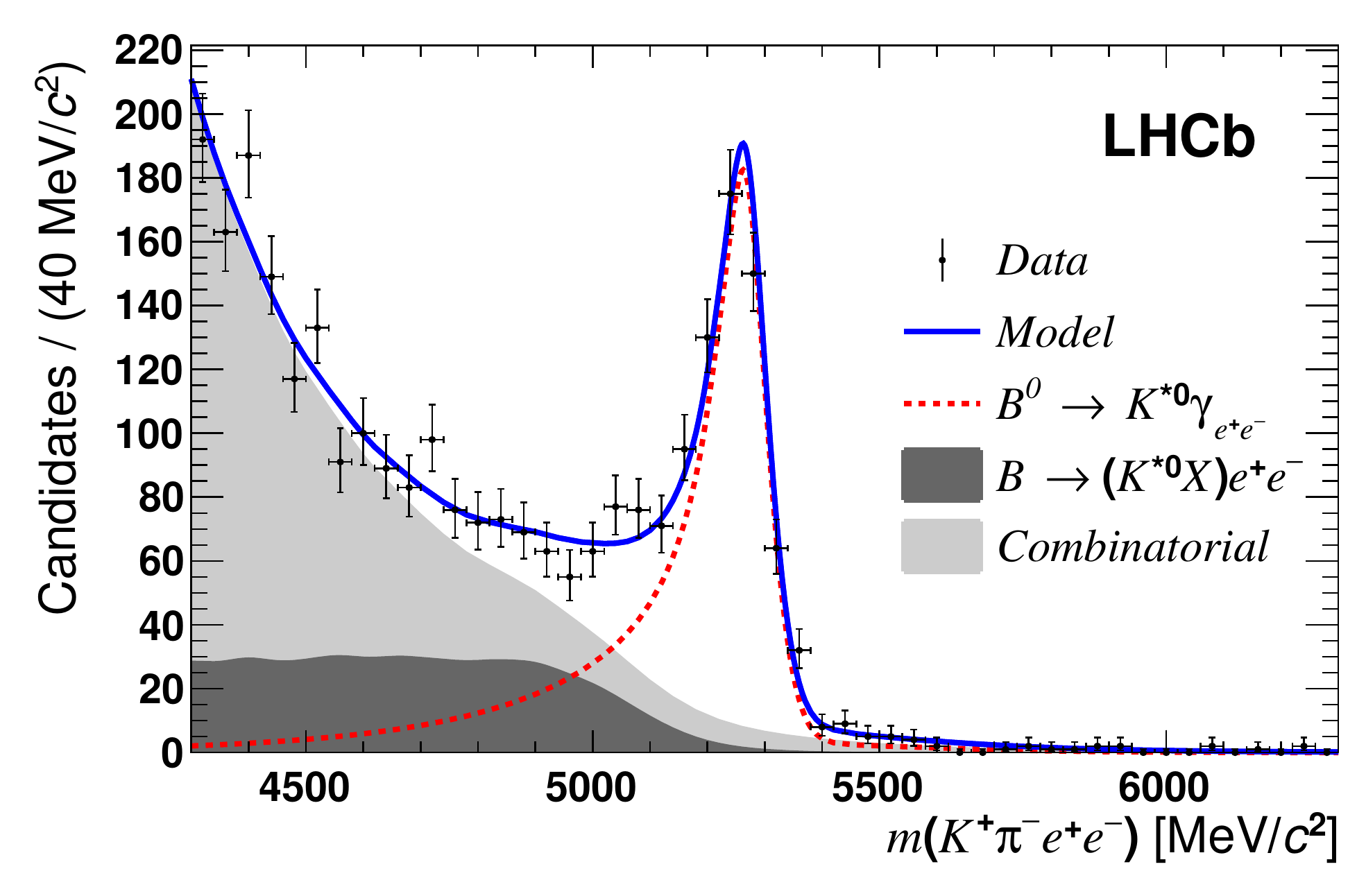}\put(-175,123){(a)}
    \includegraphics[angle=0,width=0.495\textwidth]{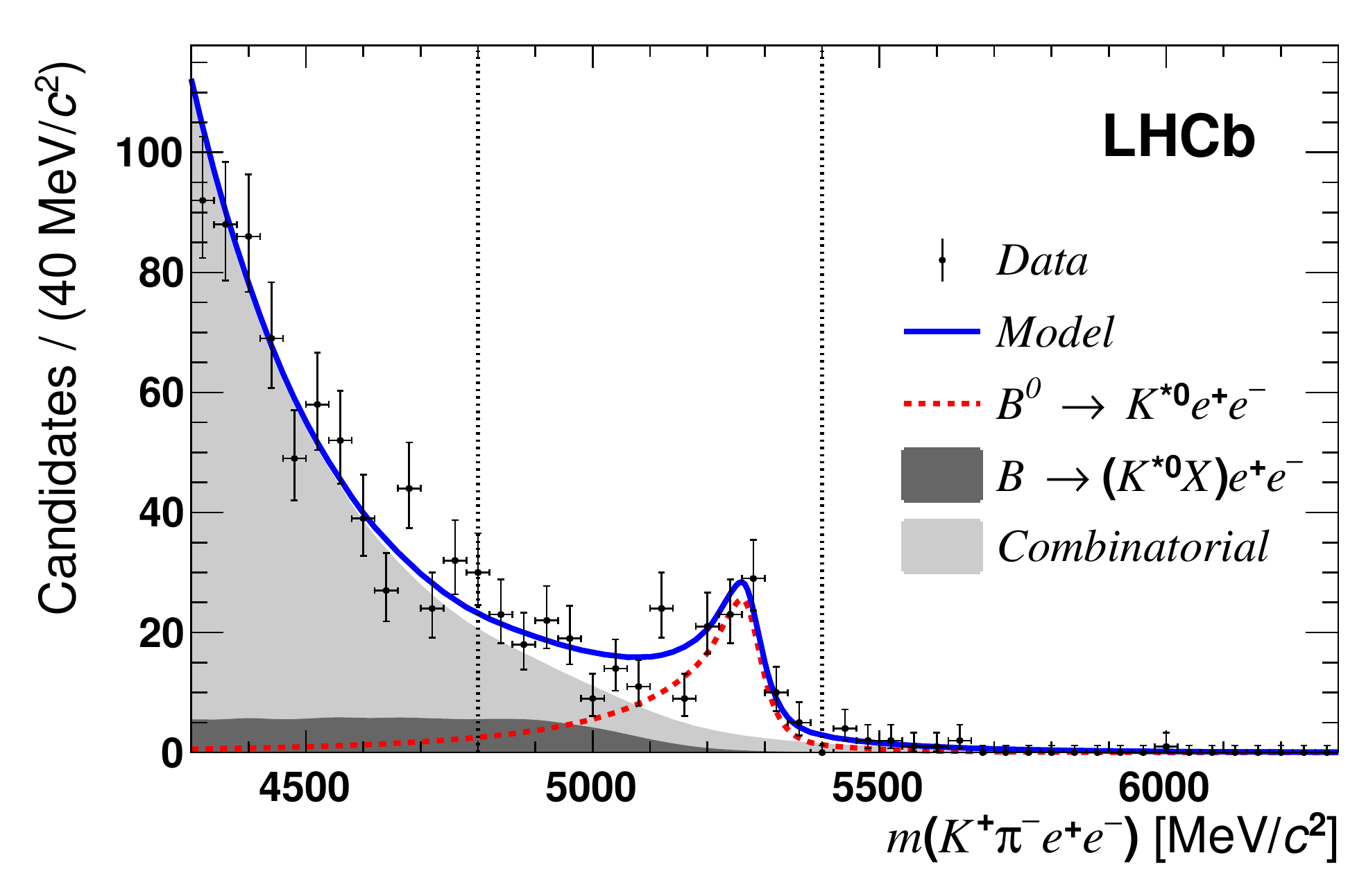}\put(-175,123){(b)}
  \caption{\small Invariant mass distribution for (a) the \BdKstGamToee and (b) the \BdKstee decay modes and the three trigger categories grouped together. The dashed line is the signal PDF, the light grey area corresponds to the combinatorial background and the dark grey area is the PR background. The solid line is the total PDF. The two vertical dotted lines on the \BdKstee plot indicate the signal window that is used in the angular fit.}  
  \label{fig:eeKstarMassFit}
\end{figure}

The signal shape parameters are fixed to the values obtained from fits to simulation but the widths and mean values are corrected for data simulation differences using \hbox{\BdToJPsieeKst} as a control channel. 
Since the photon pole contribution dominates in the low-\qsq region, the PR background is expected to be similar for \BdKstee and \BdKstGam. The 
large branching fraction of the decay \BdKstGam allows the fractions of PR background relative to the signal yield to be determined from the data. 
These fractions are extracted from a fit to a larger sample of events obtained by removing the requirements on the lower bound of the  \epem invariant mass and on $\sigma_z(\epem)$ and therefore dominated by \BdKstGamToee events. The invariant mass distribution, together with the PDFs resulting from this fit, is shown in Fig.~\ref{fig:eeKstarMassFit}(a) for the three trigger categories grouped together.
The corresponding distribution for the \BdKstee fit is shown in Fig.~\ref{fig:eeKstarMassFit}(b). There are  $150 \pm 17$ \BdToeeKst signal events, $106 \pm 16$ PR background events and $681 \pm 32$ combinatorial background events in the 4300$\--$6300\mevcc  window. 

In this wide mass window, the sample is dominated by combinatorial background, whose angular shape is difficult to model. Furthermore the angular distributions depend on the kinematic properties of the background and may thus vary as functions of mass. Hence, the angular fit is performed in a narrower mass window from 4800\mevcc to 5400\mevcc.
In this restricted window there are 124 \BdToeeKst signal events, 38 PR and 83 combinatorial background events,  corresponding to a signal-to-background ratio of the order of one. About half of these events belong to the electron hardware trigger category and the rest are equally distributed between the other two categories.

%% file: Angles_Acceptance_Backgrounds.tex
\section{Angular acceptance and angular modelling of the backgrounds}
\label{sec:acceptance}
\subsection{Angular acceptance} 
The angular acceptance is factorised as $\varepsilon (\ctl,\ctk,\phit) = \varepsilon(\ctl)\varepsilon(\ctk)\varepsilon(\phit)$ as supported by simulation studies. 
 The three corresponding one-dimensional angular distributions for the \BdToeeKst decay are distorted by the  geometrical acceptance of the detector, the trigger, the event reconstruction and the selection. 
Furthermore, their precise shapes depend upon the various trigger categories, each being enriched in events with different kinematic properties. For the \phit angle, a uniform acceptance is expected. However, there are distortions in both the \ctl and \ctk distributions, mainly arising from requirements on the transverse  momenta of the particles. The \ctk acceptance is asymmetric due to the momentum imbalance between the kaon and the pion from the \Kstarz decay in the laboratory frame due to their different masses. The \ctk and \ctl acceptance distributions are modelled on simulated \BdToeeKst events with Legendre polynomials of fourth order. The
functions chosen to model the \ctl acceptance are assumed to be symmetric and modified by a linear term to estimate the systematic uncertainty on the \ATRe parameter. For the \phit acceptance, no significant deviation from uniformity is observed. To estimate the systematic uncertainty, modulations in $\cos 2 \phit$ or $\sin 2 \phit$ are allowed. Such modulations are the most harmful ones since they may be confused with physics processes yielding non-zero values of \ATD or \ATIm. 

\subsection{Angular modelling of the backgrounds} 
In the mass window $4800< \meeKstar <5400\mevcc$ used in the angular analysis, about one third of the events are combinatorial background. The 
angular distribution of these events is described by the product of three independent distributions for \ctl , \ctk and \phit.
This background largely dominates at low \meeKstar: between 4300\mevcc and 4800\mevcc, 
about 90\% of the events are combinatorial background according to the mass fit shown in Fig.~\ref{fig:eeKstarMassFit}. However, the angular distributions of the background depend upon \meeKstar\ and  the information from the lower mass window cannot be used directly for modelling the signal region. The effect of this correlation is extracted from a sample of data events
selected with a looser BDT requirement but excluding the region of the BDT response corresponding to the signal. With this selection the sample 
is dominated by background in the whole mass range. The \ctk background distributions are modelled as first order polynomials. The \ctl background distributions are modelled with polynomial functions with third and fourth order terms. The \phit distributions are compatible with being uniform. This method assumes that there is no strong correlation between the BDT response and \meeKstar. This assumption is tested by subdividing the sample of events with looser BDT response and comparing the differences between the angular shapes predicted by this procedure and those observed. These differences are smaller than the statistical uncertainties of the parameters used to describe the angular shapes. The statistical uncertainties are thus used to assess the size of the systematic uncertainties due to the combinatorial background modelling. 

\par
The PR background accounts for about 15\% of the events in the angular fit mass window. These events cannot be treated in the same way as the combinatorial ones. 
Since only one or two particles are not reconstructed, the observed angular distributions retain some of the features induced by the dynamics of the decay.
Hence, they are modelled using the same functional shapes as the signal, but with independent physics parameters, \FLPH, \ATDPH, \ATImPH and \ATRePH . The loss of one or more final-state hadrons leads to a smaller apparent polarisation of the \Kstarz . While on \BdKstGam simulated events the \FL parameter is found to be zero, it reaches 17\% for simulated $\decay{B}{\g\kaon^{**}(\decay{}{\kaon\pion X}})$ events. 
Since in the SM one expects an \FL value of the order of 15 to 20\%, \FLPH is assumed to be equal to $1/3$, which is equivalent to no polarisation. This parameter is varied between 17\%  and 50\% to assess the size of the systematic uncertainty associated with this hypothesis.
Similarly, the loss of information due to the unreconstructed particles leads to a damping of the transverse asymmetries of the PR background, \ATDPH, \ATImPH and \ATRePH, compared to those of the signal. The signal transverse asymmetries are expected to be small in the SM, therefore their values are set to zero to describe the angular shape of the PR background. For \ATDPH and \ATImPH the validity of this assumption is tested by comparing
angular fits to  $\decay{B}{\jpsi\kaon^{**}(\decay{}{\kaon\pion X}})$ and $\decay{\Bz}{\jpsi\Kstarz}$ simulated events, which confirms a damping factor compatible with zero. 
The systematic uncertainty associated with this assumption is estimated by varying \ATDPH and \ATImPH up to half of the fitted signal values of \ATD and \ATIm, \ie assuming a damping factor of 0.5. 
For the \ATRePH parameter, however, one cannot estimate a damping factor with the same method since in the $\decay{B}{\jpsi\Kstarz}$ decay the value of $\ATRe$ is zero. The systematic uncertainty is evaluated by allowing the \ATRePH parameter to be as high as the \ATRe  value obtained from the \BdToeeKst angular fit.

%% file: FullFit.tex
\section{Measurement of the angular observables}
\label{sec:fitRes}
\subsection{Fit results}
\begin{figure}[tbp]
  \centering
  \includegraphics[angle=0,width=0.495\textwidth]{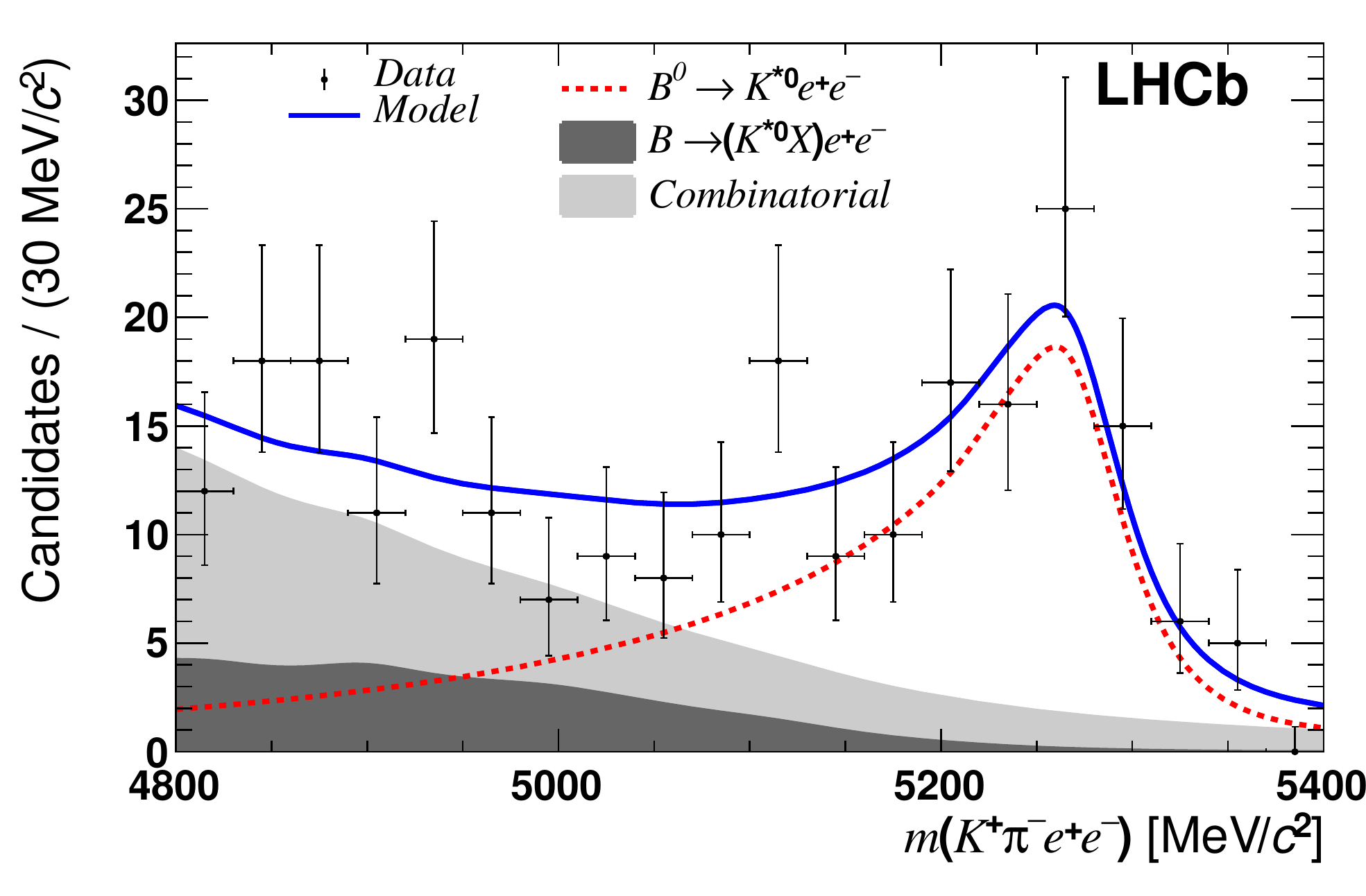}
  \includegraphics[angle=0,width=0.495\textwidth]{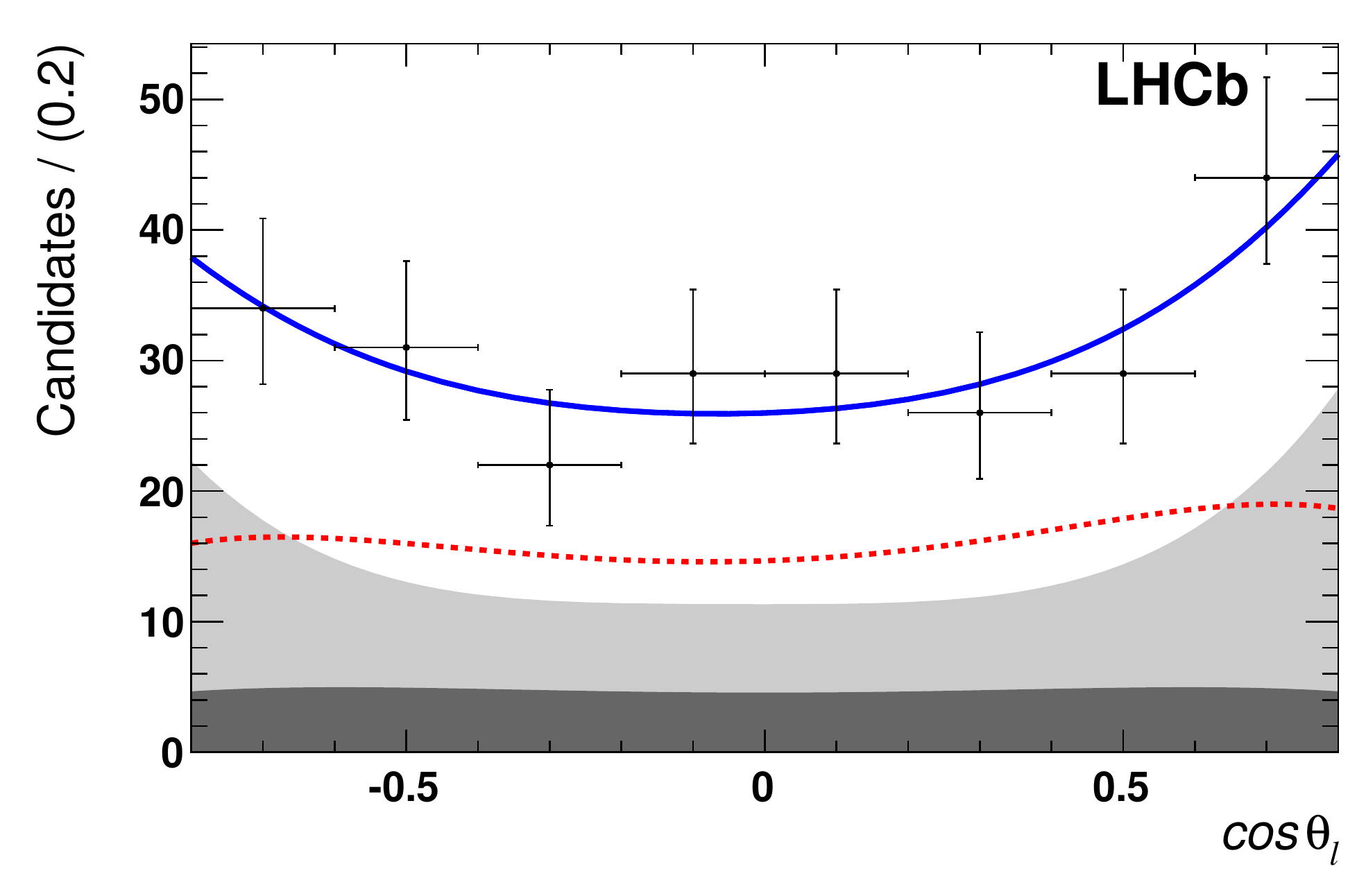}
  \includegraphics[angle=0,width=0.495\textwidth]{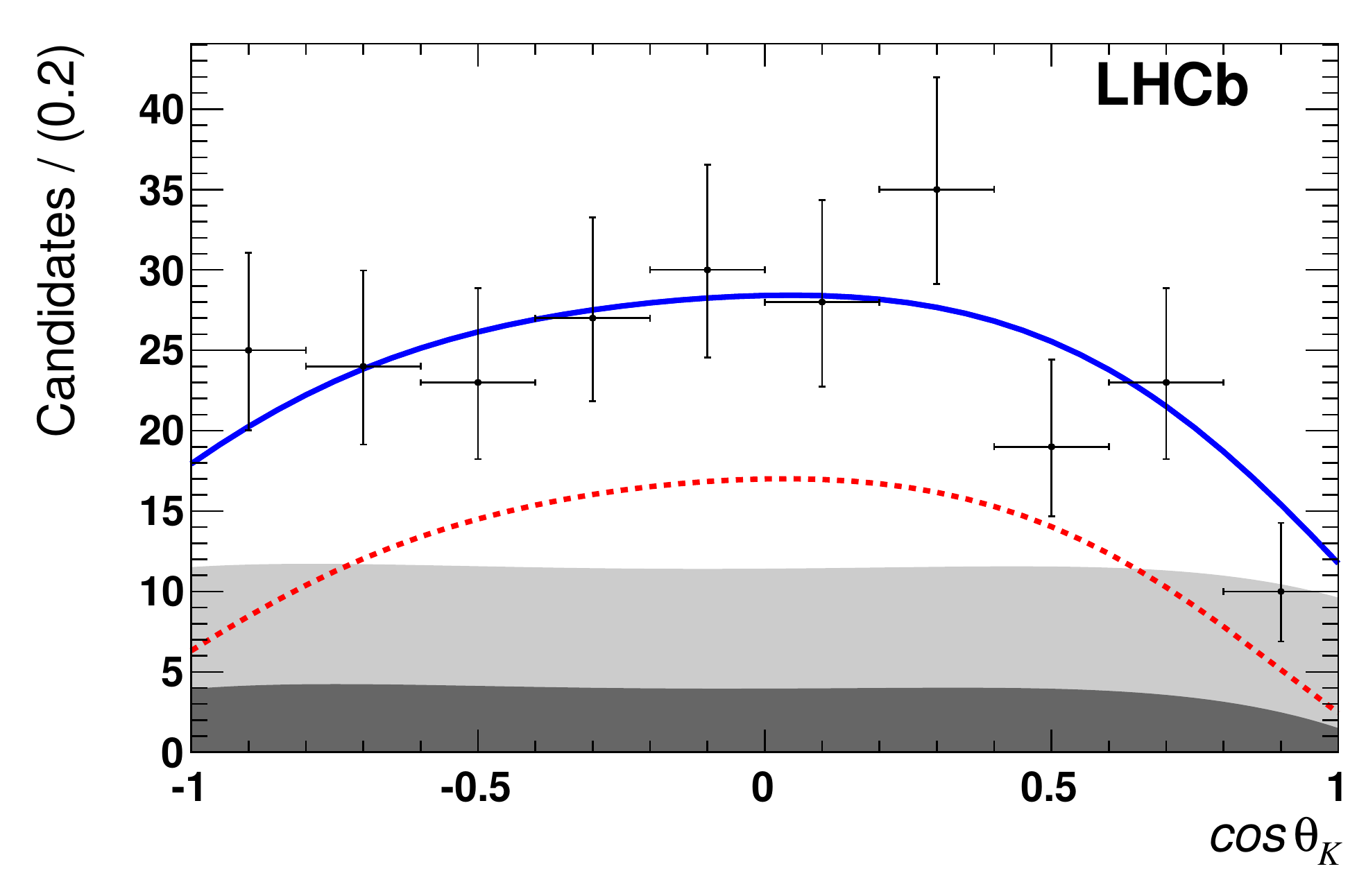}
  \includegraphics[angle=0,width=0.495\textwidth]{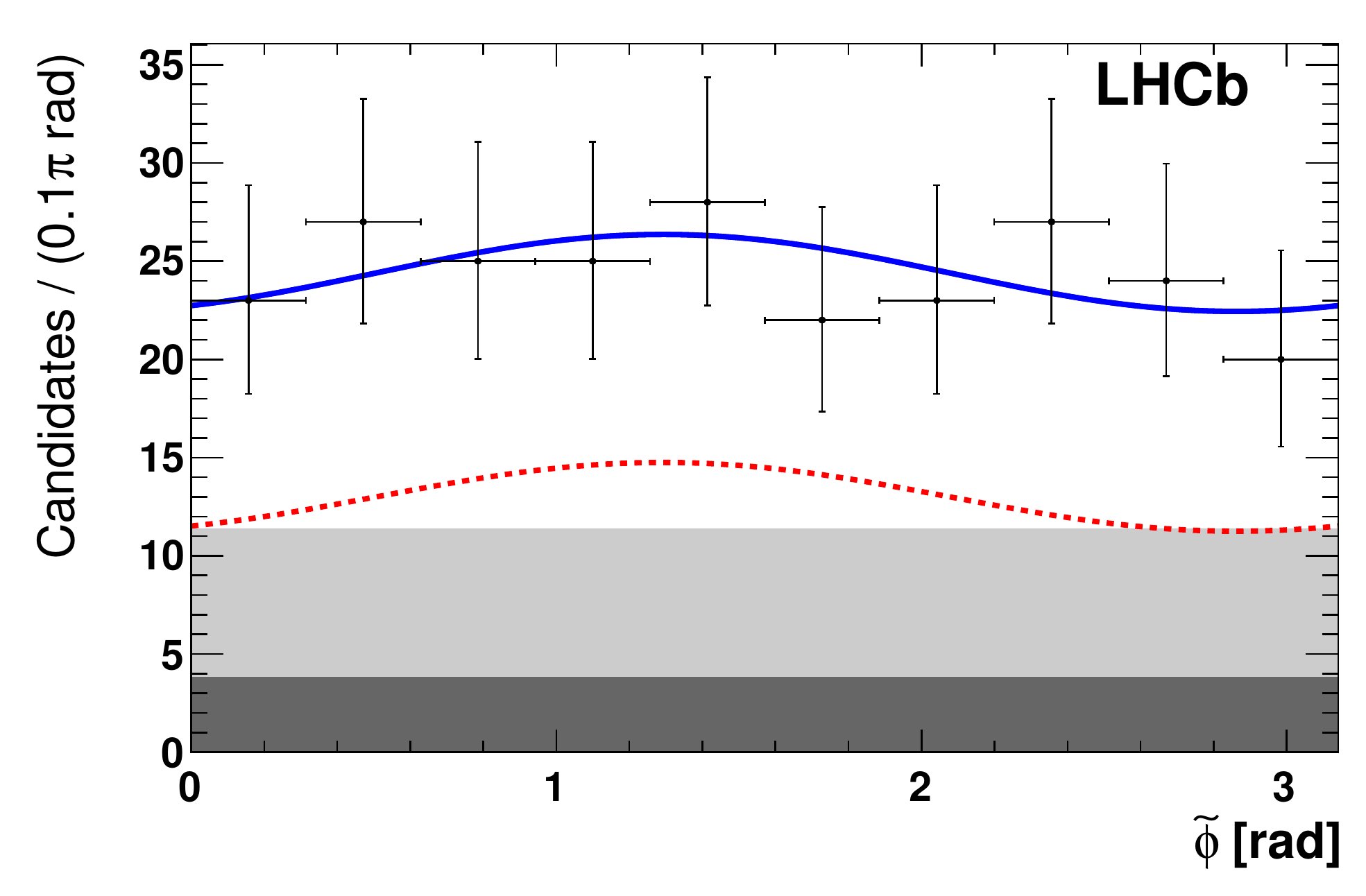}
 \caption{\small Distributions of the $K^+\pi^- \epem$ invariant mass, \ctl, \ctk and \phit variables for the  \BdKstee decay mode and the three trigger categories grouped together. The dashed line is the signal PDF, the light grey area corresponds to the combinatorial background, the dark grey area is the PR background. The solid line is the total PDF.}  
 \label{fig:DataFitResult}
\end{figure}

To measure the four angular observables, \FL, \ATD, \ATIm and \ATRe, an unbinned maximum likelihood fit is performed on the \meeKstar, \ctl, \ctk and \phit distributions in the signal window defined in Sec.~\ref{sec:MassFit}. The inclusion of \meeKstar\ in the fit strongly improves its statistical power since the level of background varies significantly within the signal mass window. 
The fit is performed simultaneously on the three trigger categories sharing the fit parameters associated with the angular observables. The mass PDFs for the three components (signal, PR background and combinatorial background) are obtained from the fit described in Sec.~\ref{sec:MassFit}. The angular PDFs for the signal are obtained by multiplying the formula of Eq.~\ref{eq:AngDistr} by the acceptance described in 
Sec.~\ref{sec:acceptance}. Similarly, the angular PDFs for the PR background are modelled by using Eq.~\ref{eq:AngDistr} and the acceptance described in Sec.~\ref{sec:acceptance} and setting \mbox{\FLPH = 0.33} and \mbox{\ATDPH = \ATImPH = \ATRePH = 0}. Finally, the angular PDFs for the combinatorial background are described in Sec.~\ref{sec:acceptance}. The combinatorial and PR background fractions are constrained to the
values calculated from the mass fit described in Sec.~\ref{sec:MassFit}. 
The fit is validated using a large number of pseudo-experiments that include all the components of the fits. Several input values for the angular observables, \FL, \ATD, \ATIm and \ATRe, are studied including those associated with NP models, and the fit results are in good agreement with the inputs. The fitting procedure is also verified using a large sample of fully simulated events; the fitted values of  \FL, \ATD, \ATIm and \ATRe are in excellent agreement with the generated ones. This validates not only the fit but also the assumption that the angular acceptance factorises. 
The distributions of \meeKstar, \ctl, \ctk and \phit, together with the likelihood projections resulting from the fit, are shown in Fig.~\ref{fig:DataFitResult} and the fit results are given in Table~\ref{tab:Results}. The fitted values of  \FL, \ATD, \ATIm and \ATRe are corrected for the $(3.8 \pm 1.9)$\% contamination from \BdKstGamToee decays, assuming that \FLKG, \ATDKG, \ATImKG and  \ATReKG are all equal to zero, and are used for the computation of the systematic uncertainties related to the angular
description of the PR background. The fitted values are also corrected for the small fit biases due to the limited size of the data sample.
\begin{table}[tbp]
\centering
\caption{Fit results for the angular observables \FL, \ATD, \ATIm and \ATRe. The second column corresponds to the uncorrected values directly obtained from the fit while the third column gives the final results after the correction for the $(3.8 \pm 1.9)$\% of \BdKstGamToee contamination and for the small fit biases due to the limited size of the data sample. The first uncertainty is statistical and the second systematic.}
\begin{tabular}{|l|c|c|}
\hline
 & Uncorrected values & Corrected values \\ \hline
 \FL  &  $\phantom{+}0.15 \pm 0.06$  & $\phantom{+}0.16 \pm 0.06 \pm 0.03$  \\
 \ATD & $-0.22 \pm 0.23$	 & $-0.23 \pm 0.23 \pm 0.05 $ \\
 \ATIm & $+0.14 \pm 0.22$	 	&$+0.14 \pm 0.22 \pm 0.05$	  \\
 \ATRe &  $+0.09 \pm 0.18 $ & $+0.10 \pm 0.18 \pm 0.05 $ \\
 \hline
\end{tabular}
\label{tab:Results}
\end{table}

\subsection{Systematic uncertainties}
To evaluate the contributions from the possible sources of systematic uncertainty, pseudo-experiments with modified parameters are generated and fitted with the PDFs used to fit the data. Fit results are then compared with input values to assess the size of the uncertainties. 

The systematic uncertainties due to the modelling of the angular acceptance are estimated by varying the shapes introducing functional dependences that would bias the angular observables. 

The uncertainties due to the description of the shape of the combinatorial background are obtained from the uncertainties on the parameters describing the shapes and by allowing for potential $\cos2\phit$ and $\sin 2 \phit$ modulations. 

To estimate the uncertainties due to the modelling of the PR background the \FLPH parameter is varied between 0.17 and 0.5. The systematic uncertainties related to the \ATD and \ATIm observables depend on the values of the observables themselves: their sizes are assessed by varying the damping factor up to 0.5, i.e. reducing the distortions of the \phit distribution of the PR background by a factor of two compared to the signal ones. For the \ATRe parameter, the systematic
uncertainty is estimated by varying \ATRePH up to the fitted value obtained for \BdToeeKst. 

The systematic uncertainties from the \BdKstGamToee background are due to the uncertainty on the size of the contamination. 

Finally, to estimate possible biases due to the fitting procedure, a large number of pseudo-experiments are generated with the number of events observed in data and are fitted with the default PDFs. While the \ATD and \ATIm estimates are not biased, the \FL and \ATRe observables exhibit small biases (less than 10\% of the statistical uncertainties) due to the limited size of the data sample and are corrected accordingly. The values of the corrections are assigned as uncertainties (labelled as ``Fit bias'' in Table~\ref{tab:syst}) .

The systematic uncertainties are summarised in Table~\ref{tab:syst}. The systematic uncertainties on the \FL, \ATD, \ATIm and \ATRe angular observables in Table~\ref{tab:Results} are obtained by adding these contributions in quadrature. They are, in all cases, smaller than the statistical uncertainties. 

\begin{table}[tpbp]
\centering
     \caption{Summary of the systematic uncertainties.}
\begin{tabular}{|l|c|c|c|c|}
\hline
\multicolumn{1}{|c|}{Source}     & \multicolumn{1}{|c|}{$\sigma(\FL)$} & \multicolumn{1}{|c|}{$\sigma(\ATD)$} & \multicolumn{1}{|c|}{$\sigma(\ATIm)$} & \multicolumn{1}{|c|}{$\sigma(\ATRe)$} \\\hline
    Acceptance modelling        & $0.013$ & $0.038$ & $0.035$ & $0.031$ \\
Combinatorial background  & $0.006$ & $0.030$ & $0.029$ & $0.038$ \\
PR background & $0.019$ & $0.011$ & $0.007$ &  $0.009$\\    
\BdKstGam contamination & $0.003$ & $0.004$  & $0.003$ & $0.002$ \\
Fit bias & $0.008$ & \-- & \-- & 0.010 \\   \hline
Total systematic uncertainty & 0.03 & 0.05 & 0.05 & 0.05 \\ \hline
Statistical uncertainty & 0.06 & 0.23 & 0.22 & 0.18 \\ \hline
      \end{tabular}
   \label{tab:syst}
\end{table}

\subsection{Effective \qsq range of the selected \BdToeeKst signal events }
\begin{figure}[tbp]
  \centering
   \includegraphics[angle=0,width=0.6\textwidth]{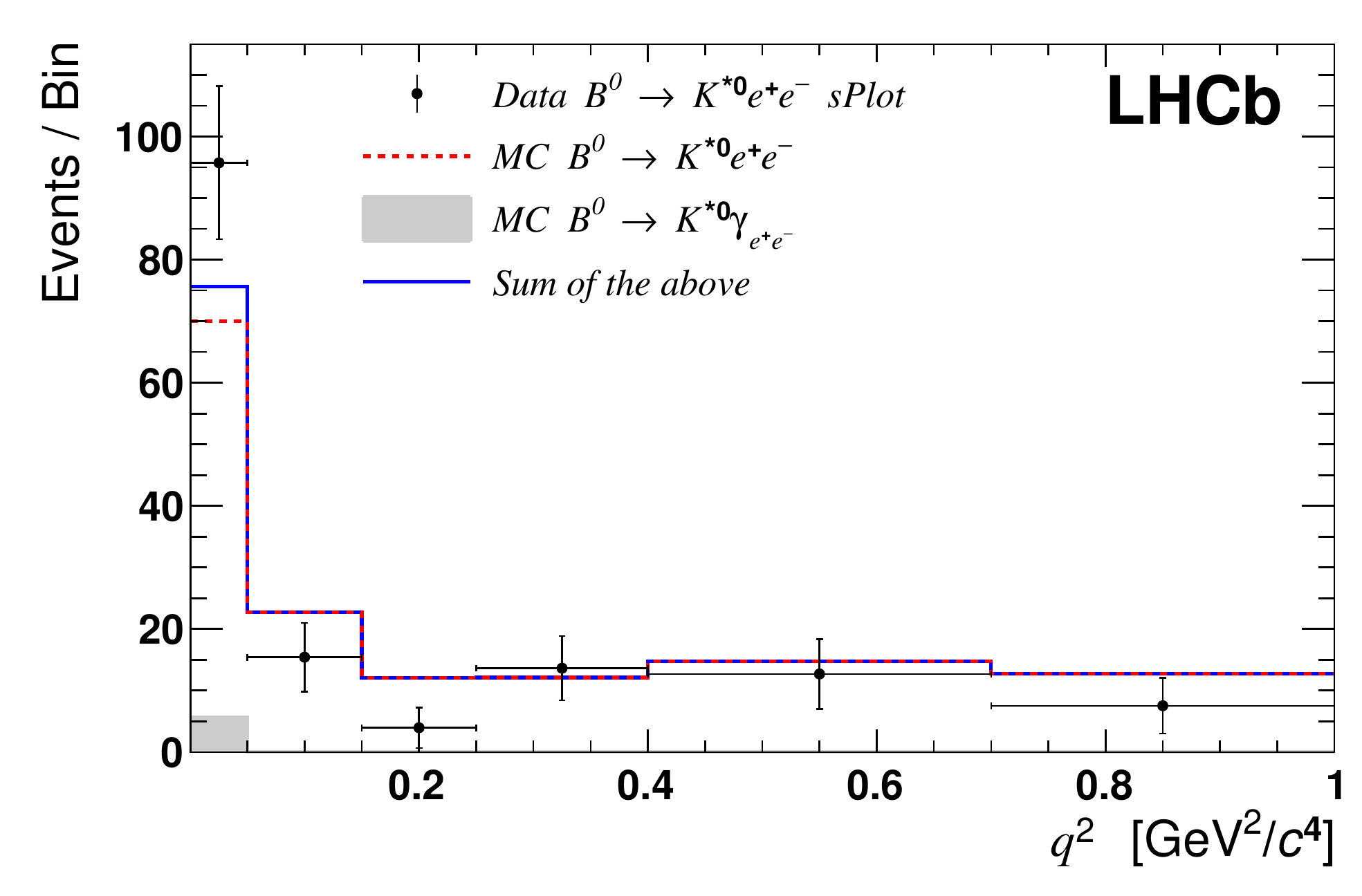}
  \caption{Distribution of the reconstructed \qsq  from an \sPlot of data (black points). The dashed line represents the \BdToeeKst contribution and the grey area corresponds to the $3.8\%$ \BdKstGamToee contamination. The solid line is the sum of the two. }
  \label{fig:sPlotq2}
\end{figure}

The distribution of the reconstructed \qsq for the signal is obtained using the \sPlot technique~\cite{Pivk:2004ty} based on the \Bz invariant mass spectrum and shown in Fig.~\ref{fig:sPlotq2}.  
Taking into account the effect of event migration in and out the \qsq bin, the average value of the true \qsq of the selected signal events is equal to $\qsq = 0.17 \pm 0.04 $\gevgevcccc.  
The acceptance as a function of the true \qsq, obtained from the \lhcb simulation, is uniform in a large domain except close to the limits of the reconstructed \qsq, 0.0004 and 1\gevgevcccc. 
Due to reconstruction effects, the  \qsq effective limits are slightly different. Because of reduced acceptance in the low-\qsq region, the value of the  lower  \qsq effective limit is increased; because of bremsstrahlung radiation, events with a true \qsq greater than 1\gevgevcccc are accepted by the selection and the  higher  \qsq effective limit is also increased. The values of these effective boundaries are obtained by requiring that in the low- and high- \qsq regions the
same number of events are obtained in a uniform acceptance model and in the \lhcb simulation. The  true \qsq effective region is thus determined to be between 0.002 and 1.12\gevgevcccc. It is checked, using the \lhcb simulation, that the average values of the true \qsq and of the angular observables evaluated with a uniform acceptance in the region between 0.002 and 1.12\gevgevcccc are in agreement with those obtained from the angular fit performed on the events selected in the
reconstructed \qsq interval 0.0004 to 1\gevgevcccc. An uncertainty on the \qsq effective limits is assigned as half of the \qsq limit modification. The  true  \qsq effective range is thus from $0.0020 \pm 0.0008$ to $1.120 \pm 0.060$\gevgevcccc. This range should be used to compare the \FL, \ATD, \ATIm and \ATRe measurements with predictions.

%% file: Summary.tex
\section{Summary}
An angular analysis of the \BdToeeKst decay is performed using proton-proton collision data, corresponding to an integrated luminosity of 3.0\invfb, collected by the \lhcb experiment in 2011 and 2012. 
Angular observables are measured for the first time in an effective \qsq range from $0.0020 \pm 0.0008$ to $1.120 \pm 0.060$\gevgevcccc. The results are
 \begin{eqnarray} \nonumber
  \FL &=&  \phantom{+}0.16 \pm 0.06 \pm 0.03 \\\nonumber
  \label{eq:DataFitResultCorr}
  \ATD &=& -0.23 \pm 0.23 \pm 0.05 \\\nonumber
   \ATIm &=& +0.14 \pm 0.22 \pm 0.05\\\nonumber
  \ATRe &=& +0.10 \pm 0.18 \pm 0.05 , \nonumber
\end{eqnarray}
where the first contribution to the uncertainty is statistical and the second systematic. The results are consistent with SM predictions~\cite{Becirevic:2011bp,Jager:2014rwa}.
For the low average value of \qsq of this analysis, the formulae  relating \ATD and \ATIm and \C7 and $\mathcal{C}_7^{'}$ in Eq.~\ref{eq:qSqToZero} are accurate at the 5\% level, for SM values of the ratios of Wilson coefficients $\mathcal{C}_{9}/\mathcal{C}_{7}$ and $\mathcal{C}_{10}/\mathcal{C}_{7}$. At this level of precision and for SM values of \C7, the ratio 
$\mathcal{C}_7^{'}/\mathcal{C}_7$ is compatible with zero. This determination is more precise than that obtained from the average of the time-dependent measurements of \CP asymmetry in \BdToKstPizGamma decays~\cite{Aubert:2008gy,Ushiroda:2006fi}.

%% file: acknowledgements.tex
\section*{Acknowledgements}
\noindent We express our gratitude to our colleagues in the CERN
accelerator departments for the excellent performance of the LHC. We
thank the technical and administrative staff at the LHCb
institutes. We acknowledge support from CERN and from the national
agencies: CAPES, CNPq, FAPERJ and FINEP (Brazil); NSFC (China);
CNRS/IN2P3 (France); BMBF, DFG, HGF and MPG (Germany); INFN (Italy); 
FOM and NWO (The Netherlands); MNiSW and NCN (Poland); MEN/IFA (Romania); 
MinES and FANO (Russia); MinECo (Spain); SNSF and SER (Switzerland); 
NASU (Ukraine); STFC (United Kingdom); NSF (USA).
The Tier1 computing centres are supported by IN2P3 (France), KIT and BMBF 
(Germany), INFN (Italy), NWO and SURF (The Netherlands), PIC (Spain), GridPP 
(United Kingdom).
We are indebted to the communities behind the multiple open 
source software packages on which we depend. We are also thankful for the 
computing resources and the access to software R\&D tools provided by Yandex LLC (Russia).
Individual groups or members have received support from 
EPLANET, Marie Sk\l{}odowska-Curie Actions and ERC (European Union), 
Conseil g\'{e}n\'{e}ral de Haute-Savoie, Labex ENIGMASS and OCEVU, 
R\'{e}gion Auvergne (France), RFBR (Russia), XuntaGal and GENCAT (Spain), Royal Society and Royal
Commission for the Exhibition of 1851 (United Kingdom).

%% file: LHCb_HD_authorlist_2014-11-18.tex
\centerline{\large\bf LHCb collaboration}
\begin{flushleft}
\small
R.~Aaij$^{41}$, 
B.~Adeva$^{37}$, 
M.~Adinolfi$^{46}$, 
A.~Affolder$^{52}$, 
Z.~Ajaltouni$^{5}$, 
S.~Akar$^{6}$, 
J.~Albrecht$^{9}$, 
F.~Alessio$^{38}$, 
M.~Alexander$^{51}$, 
S.~Ali$^{41}$
% ...

{\footnotesize \it
% ...
$ ^{5}$Clermont Universit\'{e}, Universit\'{e} Blaise Pascal, CNRS/IN2P3, LPC, Clermont-Ferrand, France\\
$ ^{6}$CPPM, Aix-Marseille Universit\'{e}, CNRS/IN2P3, Marseille, France\\
% ...
$ ^{9}$Fakult\"{a}t Physik, Technische Universit\"{a}t Dortmund, Dortmund, Germany\\
% ...
$ ^{37}$Universidad de Santiago de Compostela, Santiago de Compostela, Spain\\
$ ^{38}$European Organization for Nuclear Research (CERN), Geneva, Switzerland\\
% ...
$ ^{41}$Nikhef National Institute for Subatomic Physics, Amsterdam, The Netherlands\\
% ...
$ ^{46}$H.H. Wills Physics Laboratory, University of Bristol, Bristol, United Kingdom\\
% ...
$ ^{51}$School of Physics and Astronomy, University of Glasgow, Glasgow, United Kingdom\\
$ ^{52}$Oliver Lodge Laboratory, University of Liverpool, Liverpool, United Kingdom\\
% ...
}
\end{flushleft}

%% file: main.bbl
\ifx\mcitethebibliography\mciteundefinedmacro
\PackageError{LHCb.bst}{mciteplus.sty has not been loaded}
{This bibstyle requires the use of the mciteplus package.}\fi
\providecommand{\href}[2]{#2}
\begin{mcitethebibliography}{10}
\mciteSetBstSublistMode{n}
\mciteSetBstMaxWidthForm{subitem}{\alph{mcitesubitemcount})}
\mciteSetBstSublistLabelBeginEnd{\mcitemaxwidthsubitemform\space}
{\relax}{\relax}

\bibitem{Kruger:2005ep}
F.~Kruger and J.~Matias, \ifthenelse{\boolean{articletitles}}{\emph{{Probing
  new physics via the transverse amplitudes of \decay{\Bz}{\Kstarz (\ra
  K^+\pi^-)\ellp\ellm} at large recoil}},
  }{}\href{http://dx.doi.org/10.1103/PhysRevD.71.094009}{Phys.\ Rev.\
  \textbf{D71} (2005) 094009}, \href{http://arxiv.org/abs/hep-ph/0502060}{{\tt
  arXiv:hep-ph/0502060}}\relax
\mciteBstWouldAddEndPuncttrue
\mciteSetBstMidEndSepPunct{\mcitedefaultmidpunct}
{\mcitedefaultendpunct}{\mcitedefaultseppunct}\relax
\EndOfBibitem
\bibitem{Becirevic:2011bp}
D.~Becirevic and E.~Schneider, \ifthenelse{\boolean{articletitles}}{\emph{{On
  transverse asymmetries in \decay{B}{\Kstar\ellp\ellm}}},
  }{}\href{http://dx.doi.org/10.1016/j.nuclphysb.2011.09.004}{Nucl.\ Phys.\
  \textbf{B854} (2012) 321}, \href{http://arxiv.org/abs/1106.3283}{{\tt
  arXiv:1106.3283}}\relax
\mciteBstWouldAddEndPuncttrue
\mciteSetBstMidEndSepPunct{\mcitedefaultmidpunct}
{\mcitedefaultendpunct}{\mcitedefaultseppunct}\relax
\EndOfBibitem
\bibitem{Grossman:2000rk}
Y.~Grossman and D.~Pirjol,
  \ifthenelse{\boolean{articletitles}}{\emph{{Extracting and using photon
  polarization information in radiative B decays}},
  }{}\href{http://dx.doi.org/10.1088/1126-6708/2000/06/029}{JHEP \textbf{06}
  (2000) 029}, \href{http://arxiv.org/abs/hep-ph/0005069}{{\tt
  arXiv:hep-ph/0005069}}\relax
\mciteBstWouldAddEndPuncttrue
\mciteSetBstMidEndSepPunct{\mcitedefaultmidpunct}
{\mcitedefaultendpunct}{\mcitedefaultseppunct}\relax
\EndOfBibitem
\bibitem{Jager:2012uw}
S.~J$\ddot{\rm a}$ger and J.~M. Camalich,
  \ifthenelse{\boolean{articletitles}}{\emph{{On $ B \ra V \ell \ell$ at small
  dilepton invariant mass, power corrections, and new physics}},
  }{}\href{http://dx.doi.org/10.1007/JHEP05(2013)043}{JHEP \textbf{05} (2013)
  043}, \href{http://arxiv.org/abs/1212.2263}{{\tt arXiv:1212.2263}}\relax
\mciteBstWouldAddEndPuncttrue
\mciteSetBstMidEndSepPunct{\mcitedefaultmidpunct}
{\mcitedefaultendpunct}{\mcitedefaultseppunct}\relax
\EndOfBibitem
\bibitem{Everett:2001yy}
L.~L. Everett {\em et~al.},
  \ifthenelse{\boolean{articletitles}}{\emph{{Alternative approach to
  \decay{b}{s\g} in the uMSSM}},
  }{}\href{http://dx.doi.org/10.1088/1126-6708/2002/01/022}{JHEP \textbf{01}
  (2002) 022}, \href{http://arxiv.org/abs/hep-ph/0112126}{{\tt
  arXiv:hep-ph/0112126}}\relax
\mciteBstWouldAddEndPuncttrue
\mciteSetBstMidEndSepPunct{\mcitedefaultmidpunct}
{\mcitedefaultendpunct}{\mcitedefaultseppunct}\relax
\EndOfBibitem
\bibitem{Foster:2006ze}
J.~Foster, K.-i. Okumura, and L.~Roszkowski,
  \ifthenelse{\boolean{articletitles}}{\emph{{New constraints on SUSY flavour
  mixing in light of recent measurements at the Tevatron}},
  }{}\href{http://dx.doi.org/10.1016/j.physletb.2006.09.004}{Phys.\ Lett.\
  \textbf{B641} (2006) 452}, \href{http://arxiv.org/abs/hep-ph/0604121}{{\tt
  arXiv:hep-ph/0604121}}\relax
\mciteBstWouldAddEndPuncttrue
\mciteSetBstMidEndSepPunct{\mcitedefaultmidpunct}
{\mcitedefaultendpunct}{\mcitedefaultseppunct}\relax
\EndOfBibitem
\bibitem{Lunghi:2006hc}
E.~Lunghi and J.~Matias, \ifthenelse{\boolean{articletitles}}{\emph{{Huge
  right-handed current effects in $B \ra \Kstarz (K \pi) \ellp \ellm$ in
  supersymmetry}},
  }{}\href{http://dx.doi.org/10.1088/1126-6708/2007/04/058}{JHEP \textbf{04}
  (2007) 058}, \href{http://arxiv.org/abs/hep-ph/0612166}{{\tt
  arXiv:hep-ph/0612166}}\relax
\mciteBstWouldAddEndPuncttrue
\mciteSetBstMidEndSepPunct{\mcitedefaultmidpunct}
{\mcitedefaultendpunct}{\mcitedefaultseppunct}\relax
\EndOfBibitem
\bibitem{Goto:2007ee}
T.~Goto, Y.~Okada, T.~Shindou, and M.~Tanaka,
  \ifthenelse{\boolean{articletitles}}{\emph{{Patterns of flavor signals in
  supersymmetric models}},
  }{}\href{http://dx.doi.org/10.1103/PhysRevD.77.095010}{Phys.\ Rev.\
  \textbf{D77} (2008) 095010}, \href{http://arxiv.org/abs/0711.2935}{{\tt
  arXiv:0711.2935}}\relax
\mciteBstWouldAddEndPuncttrue
\mciteSetBstMidEndSepPunct{\mcitedefaultmidpunct}
{\mcitedefaultendpunct}{\mcitedefaultseppunct}\relax
\EndOfBibitem
\bibitem{Wei:2009zv}
Belle collaboration, J.-T. Wei {\em et~al.},
  \ifthenelse{\boolean{articletitles}}{\emph{{Measurement of the differential
  branching fraction and forward-backward asymmetry for $B \to
  K^{(*)}\ell^+\ell^-$}},
  }{}\href{http://dx.doi.org/10.1103/PhysRevLett.103.171801}{Phys.\ Rev.\
  Lett.\  \textbf{103} (2009) 171801},
  \href{http://arxiv.org/abs/0904.0770}{{\tt arXiv:0904.0770}}\relax
\mciteBstWouldAddEndPuncttrue
\mciteSetBstMidEndSepPunct{\mcitedefaultmidpunct}
{\mcitedefaultendpunct}{\mcitedefaultseppunct}\relax
\EndOfBibitem
\bibitem{Aaltonen:2011ja}
CDF collaboration, T.~Aaltonen {\em et~al.},
  \ifthenelse{\boolean{articletitles}}{\emph{{Measurements of the angular
  distributions in the decays $B \to K^{(*)} \mu^+ \mu^-$ at CDF}},
  }{}\href{http://dx.doi.org/10.1103/PhysRevLett.108.081807}{Phys.\ Rev.\
  Lett.\  \textbf{108} (2012) 081807},
  \href{http://arxiv.org/abs/1108.0695}{{\tt arXiv:1108.0695}}\relax
\mciteBstWouldAddEndPuncttrue
\mciteSetBstMidEndSepPunct{\mcitedefaultmidpunct}
{\mcitedefaultendpunct}{\mcitedefaultseppunct}\relax
\EndOfBibitem
\bibitem{LHCb-PAPER-2013-019}
LHCb collaboration, R.~Aaij {\em et~al.},
  \ifthenelse{\boolean{articletitles}}{\emph{{Differential branching fraction
  and angular analysis of the decay $B^0 \to K^{*0} \mu^+\mu^-$}},
  }{}\href{http://dx.doi.org/10.1007/JHEP08(2013)131}{JHEP \textbf{08} (2013)
  131}, \href{http://arxiv.org/abs/1304.6325}{{\tt arXiv:1304.6325}}\relax
\mciteBstWouldAddEndPuncttrue
\mciteSetBstMidEndSepPunct{\mcitedefaultmidpunct}
{\mcitedefaultendpunct}{\mcitedefaultseppunct}\relax
\EndOfBibitem
\bibitem{Lefrancois:1179865}
J.~Lefran\c{c}ois and M.~H. Schune,
  \ifthenelse{\boolean{articletitles}}{\emph{{Measuring the photon polarization
  in \decay{b}{s\g} using the \BdToeeKst decay channel}}, }{} Tech. Rep.
  LHCb-PUB-2009-008. CERN-LHCb-PUB-2009-008., CERN, Geneva, Jun, 2009\relax
\mciteBstWouldAddEndPuncttrue
\mciteSetBstMidEndSepPunct{\mcitedefaultmidpunct}
{\mcitedefaultendpunct}{\mcitedefaultseppunct}\relax
\EndOfBibitem
\bibitem{LHCb-PAPER-2013-005}
LHCb collaboration, R.~Aaij {\em et~al.},
  \ifthenelse{\boolean{articletitles}}{\emph{{Measurement of the $B^0 \to
  K^{*0}e^+e^-$ branching fraction at low dilepton mass}},
  }{}\href{http://dx.doi.org/10.1007/JHEP05(2013)159}{JHEP \textbf{05} (2013)
  159}, \href{http://arxiv.org/abs/1304.3035}{{\tt arXiv:1304.3035}}\relax
\mciteBstWouldAddEndPuncttrue
\mciteSetBstMidEndSepPunct{\mcitedefaultmidpunct}
{\mcitedefaultendpunct}{\mcitedefaultseppunct}\relax
\EndOfBibitem
\bibitem{Lu:2011jm}
C.-D. Lu and W.~Wang, \ifthenelse{\boolean{articletitles}}{\emph{{Analysis of
  $B\to K^*_J (\to K \pi) \mu^+\mu^-$ in the higher kaon resonance region}},
  }{}\href{http://dx.doi.org/10.1103/PhysRevD.85.034014}{Phys.\ Rev.\
  \textbf{D85} (2012) 034014}, \href{http://arxiv.org/abs/1111.1513}{{\tt
  arXiv:1111.1513}}\relax
\mciteBstWouldAddEndPuncttrue
\mciteSetBstMidEndSepPunct{\mcitedefaultmidpunct}
{\mcitedefaultendpunct}{\mcitedefaultseppunct}\relax
\EndOfBibitem
\bibitem{Bobeth:2008ij}
C.~Bobeth, G.~Hiller, and G.~Piranishvili,
  \ifthenelse{\boolean{articletitles}}{\emph{{CP asymmetries in $\bar{B} \to
  \bar{K}^* (\to \bar{K} \pi) \bar{\ell} \ell$ and untagged $\bar{B}_s$, $B_s
  \to \phi (\to K^{+} K^-) \bar{\ell} \ell$ decays at NLO}},
  }{}\href{http://dx.doi.org/10.1088/1126-6708/2008/07/106}{JHEP \textbf{07}
  (2008) 106}, \href{http://arxiv.org/abs/0805.2525}{{\tt
  arXiv:0805.2525}}\relax
\mciteBstWouldAddEndPuncttrue
\mciteSetBstMidEndSepPunct{\mcitedefaultmidpunct}
{\mcitedefaultendpunct}{\mcitedefaultseppunct}\relax
\EndOfBibitem
\bibitem{Aubert:2008gy}
BaBar collaboration, B.~Aubert {\em et~al.},
  \ifthenelse{\boolean{articletitles}}{\emph{{Measurement of time-dependent CP
  asymmetry in $\Bz \to \KS \piz \gamma $ decays}},
  }{}\href{http://dx.doi.org/10.1103/PhysRevD.78.071102}{Phys.\ Rev.\
  \textbf{D78} (2008) 071102}, \href{http://arxiv.org/abs/0807.3103}{{\tt
  arXiv:0807.3103}}\relax
\mciteBstWouldAddEndPuncttrue
\mciteSetBstMidEndSepPunct{\mcitedefaultmidpunct}
{\mcitedefaultendpunct}{\mcitedefaultseppunct}\relax
\EndOfBibitem
\bibitem{Ushiroda:2006fi}
Belle collaboration, Y.~Ushiroda {\em et~al.},
  \ifthenelse{\boolean{articletitles}}{\emph{{Time-dependent CP asymmetries in
  $\Bz \to \KS \piz \gamma $ transitions}},
  }{}\href{http://dx.doi.org/10.1103/PhysRevD.74.111104}{Phys.\ Rev.\
  \textbf{D74} (2006) 111104}, \href{http://arxiv.org/abs/hep-ex/0608017}{{\tt
  arXiv:hep-ex/0608017}}\relax
\mciteBstWouldAddEndPuncttrue
\mciteSetBstMidEndSepPunct{\mcitedefaultmidpunct}
{\mcitedefaultendpunct}{\mcitedefaultseppunct}\relax
\EndOfBibitem
\bibitem{Alves:2008zz}
LHCb collaboration, A.~A. Alves~Jr.\ {\em et~al.},
  \ifthenelse{\boolean{articletitles}}{\emph{{The \lhcb detector at the LHC}},
  }{}\href{http://dx.doi.org/10.1088/1748-0221/3/08/S08005}{JINST \textbf{3}
  (2008) S08005}\relax
\mciteBstWouldAddEndPuncttrue
\mciteSetBstMidEndSepPunct{\mcitedefaultmidpunct}
{\mcitedefaultendpunct}{\mcitedefaultseppunct}\relax
\EndOfBibitem
\bibitem{LHCb-DP-2014-002}
LHCb collaboration, R.~Aaij {\em et~al.},
  \ifthenelse{\boolean{articletitles}}{\emph{{LHCb detector performance}},
  }{}\href{http://arxiv.org/abs/1412.6352}{{\tt arXiv:1412.6352}}\relax
\mciteBstWouldAddEndPuncttrue
\mciteSetBstMidEndSepPunct{\mcitedefaultmidpunct}
{\mcitedefaultendpunct}{\mcitedefaultseppunct}\relax
\EndOfBibitem
\bibitem{LHCb-DP-2014-001}
R.~Aaij {\em et~al.}, \ifthenelse{\boolean{articletitles}}{\emph{{Performance
  of the LHCb Vertex Locator}},
  }{}\href{http://dx.doi.org/10.1088/1748-0221/9/09/P09007}{JINST \textbf{9}
  (2014) P09007}, \href{http://arxiv.org/abs/1405.7808}{{\tt
  arXiv:1405.7808}}\relax
\mciteBstWouldAddEndPuncttrue
\mciteSetBstMidEndSepPunct{\mcitedefaultmidpunct}
{\mcitedefaultendpunct}{\mcitedefaultseppunct}\relax
\EndOfBibitem
\bibitem{LHCb-DP-2013-003}
R.~Arink {\em et~al.}, \ifthenelse{\boolean{articletitles}}{\emph{{Performance
  of the LHCb Outer Tracker}},
  }{}\href{http://dx.doi.org/10.1088/1748-0221/9/01/P01002}{JINST \textbf{9}
  (2014) P01002}, \href{http://arxiv.org/abs/1311.3893}{{\tt
  arXiv:1311.3893}}\relax
\mciteBstWouldAddEndPuncttrue
\mciteSetBstMidEndSepPunct{\mcitedefaultmidpunct}
{\mcitedefaultendpunct}{\mcitedefaultseppunct}\relax
\EndOfBibitem
\bibitem{LHCb-DP-2012-003}
M.~Adinolfi {\em et~al.},
  \ifthenelse{\boolean{articletitles}}{\emph{{Performance of the \lhcb RICH
  detector at the LHC}},
  }{}\href{http://dx.doi.org/10.1140/epjc/s10052-013-2431-9}{Eur.\ Phys.\ J.\
  \textbf{C73} (2013) 2431}, \href{http://arxiv.org/abs/1211.6759}{{\tt
  arXiv:1211.6759}}\relax
\mciteBstWouldAddEndPuncttrue
\mciteSetBstMidEndSepPunct{\mcitedefaultmidpunct}
{\mcitedefaultendpunct}{\mcitedefaultseppunct}\relax
\EndOfBibitem
\bibitem{LHCb-DP-2012-002}
A.~A. Alves~Jr.\ {\em et~al.},
  \ifthenelse{\boolean{articletitles}}{\emph{{Performance of the LHCb muon
  system}}, }{}\href{http://dx.doi.org/10.1088/1748-0221/8/02/P02022}{JINST
  \textbf{8} (2013) P02022}, \href{http://arxiv.org/abs/1211.1346}{{\tt
  arXiv:1211.1346}}\relax
\mciteBstWouldAddEndPuncttrue
\mciteSetBstMidEndSepPunct{\mcitedefaultmidpunct}
{\mcitedefaultendpunct}{\mcitedefaultseppunct}\relax
\EndOfBibitem
\bibitem{LHCb-DP-2012-004}
R.~Aaij {\em et~al.}, \ifthenelse{\boolean{articletitles}}{\emph{{The \lhcb
  trigger and its performance in 2011}},
  }{}\href{http://dx.doi.org/10.1088/1748-0221/8/04/P04022}{JINST \textbf{8}
  (2013) P04022}, \href{http://arxiv.org/abs/1211.3055}{{\tt
  arXiv:1211.3055}}\relax
\mciteBstWouldAddEndPuncttrue
\mciteSetBstMidEndSepPunct{\mcitedefaultmidpunct}
{\mcitedefaultendpunct}{\mcitedefaultseppunct}\relax
\EndOfBibitem
\bibitem{BBDT}
V.~V. Gligorov and M.~Williams,
  \ifthenelse{\boolean{articletitles}}{\emph{{Efficient, reliable and fast
  high-level triggering using a bonsai boosted decision tree}},
  }{}\href{http://dx.doi.org/10.1088/1748-0221/8/02/P02013}{JINST \textbf{8}
  (2013) P02013}, \href{http://arxiv.org/abs/1210.6861}{{\tt
  arXiv:1210.6861}}\relax
\mciteBstWouldAddEndPuncttrue
\mciteSetBstMidEndSepPunct{\mcitedefaultmidpunct}
{\mcitedefaultendpunct}{\mcitedefaultseppunct}\relax
\EndOfBibitem
\bibitem{Sjostrand:2006za}
T.~Sj\"{o}strand, S.~Mrenna, and P.~Skands,
  \ifthenelse{\boolean{articletitles}}{\emph{{PYTHIA 6.4 physics and manual}},
  }{}\href{http://dx.doi.org/10.1088/1126-6708/2006/05/026}{JHEP \textbf{05}
  (2006) 026}, \href{http://arxiv.org/abs/hep-ph/0603175}{{\tt
  arXiv:hep-ph/0603175}}\relax
\mciteBstWouldAddEndPuncttrue
\mciteSetBstMidEndSepPunct{\mcitedefaultmidpunct}
{\mcitedefaultendpunct}{\mcitedefaultseppunct}\relax
\EndOfBibitem
\bibitem{Sjostrand:2007gs}
T.~Sj\"{o}strand, S.~Mrenna, and P.~Skands,
  \ifthenelse{\boolean{articletitles}}{\emph{{A brief introduction to PYTHIA
  8.1}}, }{}\href{http://dx.doi.org/10.1016/j.cpc.2008.01.036}{Comput.\ Phys.\
  Commun.\  \textbf{178} (2008) 852},
  \href{http://arxiv.org/abs/0710.3820}{{\tt arXiv:0710.3820}}\relax
\mciteBstWouldAddEndPuncttrue
\mciteSetBstMidEndSepPunct{\mcitedefaultmidpunct}
{\mcitedefaultendpunct}{\mcitedefaultseppunct}\relax
\EndOfBibitem
\bibitem{LHCb-PROC-2010-056}
I.~Belyaev {\em et~al.}, \ifthenelse{\boolean{articletitles}}{\emph{{Handling
  of the generation of primary events in Gauss, the LHCb simulation
  framework}}, }{}\href{http://dx.doi.org/10.1109/NSSMIC.2010.5873949}{Nuclear
  Science Symposium Conference Record (NSS/MIC) \textbf{IEEE} (2010)
  1155}\relax
\mciteBstWouldAddEndPuncttrue
\mciteSetBstMidEndSepPunct{\mcitedefaultmidpunct}
{\mcitedefaultendpunct}{\mcitedefaultseppunct}\relax
\EndOfBibitem
\bibitem{Lange:2001uf}
D.~J. Lange, \ifthenelse{\boolean{articletitles}}{\emph{{The EvtGen particle
  decay simulation package}},
  }{}\href{http://dx.doi.org/10.1016/S0168-9002(01)00089-4}{Nucl.\ Instrum.\
  Meth.\  \textbf{A462} (2001) 152}\relax
\mciteBstWouldAddEndPuncttrue
\mciteSetBstMidEndSepPunct{\mcitedefaultmidpunct}
{\mcitedefaultendpunct}{\mcitedefaultseppunct}\relax
\EndOfBibitem
\bibitem{Golonka:2005pn}
P.~Golonka and Z.~Was, \ifthenelse{\boolean{articletitles}}{\emph{{PHOTOS Monte
  Carlo: A precision tool for QED corrections in $Z$ and $W$ decays}},
  }{}\href{http://dx.doi.org/10.1140/epjc/s2005-02396-4}{Eur.\ Phys.\ J.\
  \textbf{C45} (2006) 97}, \href{http://arxiv.org/abs/hep-ph/0506026}{{\tt
  arXiv:hep-ph/0506026}}\relax
\mciteBstWouldAddEndPuncttrue
\mciteSetBstMidEndSepPunct{\mcitedefaultmidpunct}
{\mcitedefaultendpunct}{\mcitedefaultseppunct}\relax
\EndOfBibitem
\bibitem{Allison:2006ve}
Geant4 collaboration, J.~Allison {\em et~al.},
  \ifthenelse{\boolean{articletitles}}{\emph{{Geant4 developments and
  applications}}, }{}\href{http://dx.doi.org/10.1109/TNS.2006.869826}{IEEE
  Trans.\ Nucl.\ Sci.\  \textbf{53} (2006) 270}\relax
\mciteBstWouldAddEndPuncttrue
\mciteSetBstMidEndSepPunct{\mcitedefaultmidpunct}
{\mcitedefaultendpunct}{\mcitedefaultseppunct}\relax
\EndOfBibitem
\bibitem{Agostinelli:2002hh}
Geant4 collaboration, S.~Agostinelli {\em et~al.},
  \ifthenelse{\boolean{articletitles}}{\emph{{Geant4: A simulation toolkit}},
  }{}\href{http://dx.doi.org/10.1016/S0168-9002(03)01368-8}{Nucl.\ Instrum.\
  Meth.\  \textbf{A506} (2003) 250}\relax
\mciteBstWouldAddEndPuncttrue
\mciteSetBstMidEndSepPunct{\mcitedefaultmidpunct}
{\mcitedefaultendpunct}{\mcitedefaultseppunct}\relax
\EndOfBibitem
\bibitem{LHCb-PROC-2011-006}
M.~Clemencic {\em et~al.}, \ifthenelse{\boolean{articletitles}}{\emph{{The
  \lhcb simulation application, Gauss: Design, evolution and experience}},
  }{}\href{http://dx.doi.org/10.1088/1742-6596/331/3/032023}{{J.\ Phys.\ Conf.\
  Ser.\ } \textbf{331} (2011) 032023}\relax
\mciteBstWouldAddEndPuncttrue
\mciteSetBstMidEndSepPunct{\mcitedefaultmidpunct}
{\mcitedefaultendpunct}{\mcitedefaultseppunct}\relax
\EndOfBibitem
\bibitem{Breiman}
L.~Breiman, J.~H. Friedman, R.~A. Olshen, and C.~J. Stone, {\em Classification
  and regression trees}, Wadsworth international group, Belmont, California,
  USA, 1984\relax
\mciteBstWouldAddEndPuncttrue
\mciteSetBstMidEndSepPunct{\mcitedefaultmidpunct}
{\mcitedefaultendpunct}{\mcitedefaultseppunct}\relax
\EndOfBibitem
\bibitem{AdaBoost}
R.~E. Schapire and Y.~Freund, \ifthenelse{\boolean{articletitles}}{\emph{A
  decision-theoretic generalization of on-line learning and an application to
  boosting}, }{}\href{http://dx.doi.org/10.1006/jcss.1997.1504}{Jour.\ Comp.\
  and Syst.\ Sc.\  \textbf{55} (1997) 119}\relax
\mciteBstWouldAddEndPuncttrue
\mciteSetBstMidEndSepPunct{\mcitedefaultmidpunct}
{\mcitedefaultendpunct}{\mcitedefaultseppunct}\relax
\EndOfBibitem
\bibitem{Abulencia:2005pw}
CDF collaboration, A.~Abulencia {\em et~al.},
  \ifthenelse{\boolean{articletitles}}{\emph{{Search for $B_s \to \mu^+ \mu^-$
  and $B_d \to \mu^+ \mu^-$ decays in $p\bar{p}$ collisions with CDF II}},
  }{}\href{http://dx.doi.org/10.1103/PhysRevLett.95.221805}{Phys.\ Rev.\ Lett.\
   \textbf{95} (2005) 221805}, \href{http://arxiv.org/abs/hep-ex/0508036}{{\tt
  arXiv:hep-ex/0508036}}\relax
\mciteBstWouldAddEndPuncttrue
\mciteSetBstMidEndSepPunct{\mcitedefaultmidpunct}
{\mcitedefaultendpunct}{\mcitedefaultseppunct}\relax
\EndOfBibitem
\bibitem{Borsellino:1953zz}
A.~Borsellino, \ifthenelse{\boolean{articletitles}}{\emph{{Momentum transfer
  and angle of divergence of pairs produced by photons}},
  }{}\href{http://dx.doi.org/10.1103/PhysRev.89.1023}{Phys.\ Rev.\  \textbf{89}
  (1953) 1023}\relax
\mciteBstWouldAddEndPuncttrue
\mciteSetBstMidEndSepPunct{\mcitedefaultmidpunct}
{\mcitedefaultendpunct}{\mcitedefaultseppunct}\relax
\EndOfBibitem
\bibitem{Korchin:2010uc}
A.~Y. Korchin and V.~A. Kovalchuk,
  \ifthenelse{\boolean{articletitles}}{\emph{{Contribution of low-lying vector
  resonances to polarization observables in \BdKstee decay}},
  }{}\href{http://dx.doi.org/10.1103/PhysRevD.82.034013}{Phys.\ Rev.\
  \textbf{D82} (2010) 034013}, \href{http://arxiv.org/abs/1004.3647}{{\tt
  arXiv:1004.3647}}\relax
\mciteBstWouldAddEndPuncttrue
\mciteSetBstMidEndSepPunct{\mcitedefaultmidpunct}
{\mcitedefaultendpunct}{\mcitedefaultseppunct}\relax
\EndOfBibitem
\bibitem{LHCb-PAPER-2014-024}
LHCb collaboration, R.~Aaij {\em et~al.},
  \ifthenelse{\boolean{articletitles}}{\emph{{Test of lepton universality using
  $B^+\to K^+\ell^+\ell^-$ decays}},
  }{}\href{http://dx.doi.org/10.1103/PhysRevLett.113.151601}{Phys.\ Rev.\
  Lett.\  \textbf{113} (2014) 151601},
  \href{http://arxiv.org/abs/1406.6482}{{\tt arXiv:1406.6482}}\relax
\mciteBstWouldAddEndPuncttrue
\mciteSetBstMidEndSepPunct{\mcitedefaultmidpunct}
{\mcitedefaultendpunct}{\mcitedefaultseppunct}\relax
\EndOfBibitem
\bibitem{Skwarnicki:1986xj}
T.~Skwarnicki, {\em {A study of the radiative cascade transitions between the
  Upsilon-prime and Upsilon resonances}}, PhD thesis, Institute of Nuclear
  Physics, Krakow, 1986,
  {\href{http://inspirehep.net/record/230779/files/230779.pdf}{DESY-F31-86-02}}\relax
\mciteBstWouldAddEndPuncttrue
\mciteSetBstMidEndSepPunct{\mcitedefaultmidpunct}
{\mcitedefaultendpunct}{\mcitedefaultseppunct}\relax
\EndOfBibitem
\bibitem{Cranmer:2000du}
K.~S. Cranmer, \ifthenelse{\boolean{articletitles}}{\emph{{Kernel estimation in
  high-energy physics}},
  }{}\href{http://dx.doi.org/10.1016/S0010-4655(00)00243-5}{Comput.\ Phys.\
  Commun.\  \textbf{136} (2001) 198},
  \href{http://arxiv.org/abs/hep-ex/0011057}{{\tt arXiv:hep-ex/0011057}}\relax
\mciteBstWouldAddEndPuncttrue
\mciteSetBstMidEndSepPunct{\mcitedefaultmidpunct}
{\mcitedefaultendpunct}{\mcitedefaultseppunct}\relax
\EndOfBibitem
\bibitem{Pivk:2004ty}
M.~Pivk and F.~R. Le~Diberder,
  \ifthenelse{\boolean{articletitles}}{\emph{{sPlot: A statistical tool to
  unfold data distributions}},
  }{}\href{http://dx.doi.org/10.1016/j.nima.2005.08.106}{Nucl.\ Instrum.\
  Meth.\  \textbf{A555} (2005) 356},
  \href{http://arxiv.org/abs/physics/0402083}{{\tt
  arXiv:physics/0402083}}\relax
\mciteBstWouldAddEndPuncttrue
\mciteSetBstMidEndSepPunct{\mcitedefaultmidpunct}
{\mcitedefaultendpunct}{\mcitedefaultseppunct}\relax
\EndOfBibitem
\bibitem{Jager:2014rwa}
S.~J$\ddot{\rm a}$ger and J.~M. Camalich,
  \ifthenelse{\boolean{articletitles}}{\emph{{Reassessing the discovery
  potential of the $B \to K^{*} \ell^+\ell^-$ decays in the large-recoil
  region: SM challenges and BSM opportunities}},
  }{}\href{http://arxiv.org/abs/1412.3183}{{\tt arXiv:1412.3183}}\relax
\mciteBstWouldAddEndPuncttrue
\mciteSetBstMidEndSepPunct{\mcitedefaultmidpunct}
{\mcitedefaultendpunct}{\mcitedefaultseppunct}\relax
\EndOfBibitem
\end{mcitethebibliography}
